\documentclass[aps,nobibnotes,nofootinbib, nopacs, twocolumn,showkeys,superscriptaddress,groupedaddress]{revtex4}
\usepackage{amsmath,amsfonts}
\usepackage{color,stmaryrd,amsthm,amssymb,times,bbm,graphicx,tikz,verbatim, caption, subfig, rotating}
\usepackage{tikz-cd}
\usepackage{array}

\usepackage[bookmarks]{hyperref}

\DeclareFontFamily{U}{mathx}{\hyphenchar\font45}
\DeclareFontShape{U}{mathx}{m}{n}{
      <5> <6> <7> <8> <9> <10>
      <10.95> <12> <14.4> <17.28> <20.74> <24.88>
      mathx10
      }{}
\DeclareSymbolFont{mathx}{U}{mathx}{m}{n}
\DeclareFontSubstitution{U}{mathx}{m}{n}
\DeclareMathAccent{\widecheck}{0}{mathx}{"71}
\DeclareMathAccent{\wideparen}{0}{mathx}{"75}

\usepackage{amsmath,amsfonts,amssymb,bbm}

\newcommand{\bra}[1]{\langle #1 |}

\newcommand{\ket}[1]{| #1 \rangle}
\newcommand{\floor}[1]{\left\lfloor#1\right\rfloor}

\newcommand{\couic}[1]{}

\newcommand{\C}{\mathbb{C}}

\newcommand{\R}{\mathbb{R}}

\newcommand{\Z}{\mathbb{Z}}

\newcommand{\Id}{\operatorname{Id}}

\newcommand{\ii}{\mathrm i}
\newcommand{\diag}[2]{\left(\begin{array}{cc}#1 & 0\\0 & #2\end{array} \right)}

\newtheorem{rk}{Remark}

\def\gs{2}

\begin{document}

\title[]{Discrete Lorentz covariance for Quantum Walks and Quantum Cellular Automata}

\author{Pablo Arrighi}
\email{pablo.arrighi@imag.fr}
\affiliation{Univ. Grenoble Alpes, LIG, F-38000 Grenoble, France}
\affiliation{Universit\'e de Lyon, LIP, 46 all\'ee d'Italie, 69008 Lyon, France}  

\author{Stefano Facchini}
\email{stefano.facchini@imag.fr}
\affiliation{Univ. Grenoble Alpes, LIG, F-38000 Grenoble, France}

\author{Marcelo Forets}
\email{marcelo.forets@imag.fr}
\affiliation{Univ. Grenoble Alpes, LIG, F-38000 Grenoble, France}

\begin{abstract}

We formalize a notion of discrete Lorentz transforms for Quantum Walks (QW) and Quantum Cellular Automata (QCA), in $(1+1)$-dimensional discrete spacetime. The theory admits a diagrammatic representation in terms of a few local, circuit equivalence rules. Within this framework, we show the first-order-only covariance of the Dirac QW. We then introduce the Clock QW and the Clock QCA, and prove that they are exactly discrete Lorentz covariant. The theory also allows for non-homogeneous Lorentz transforms, between non-inertial frames.

\end{abstract}


\maketitle

\section{Introduction}
{\em Symmetries in Quantum Walks.} For the purpose of quantum simulation (on a quantum device) as envisioned by Feynman \cite{FeynmanQC}, or for the purpose of exploring the power and limits of discrete models of physics, a great deal of effort has gone into discretizing quantum physical phenomena. Most of these lead to Quantum Walk (QW) models of the phenomena. QWs are dynamics having the following features:
\begin{itemize}
\item The underlying spacetime is a discrete grid;
\item The evolution is unitary; 
\item It is causal, i.e. information propagates strictly at a bounded speed.
\item It is homogeneous, i.e. translation-invariant and time-independent.
\end{itemize}

\begin{figure*}[tpb]
\includegraphics{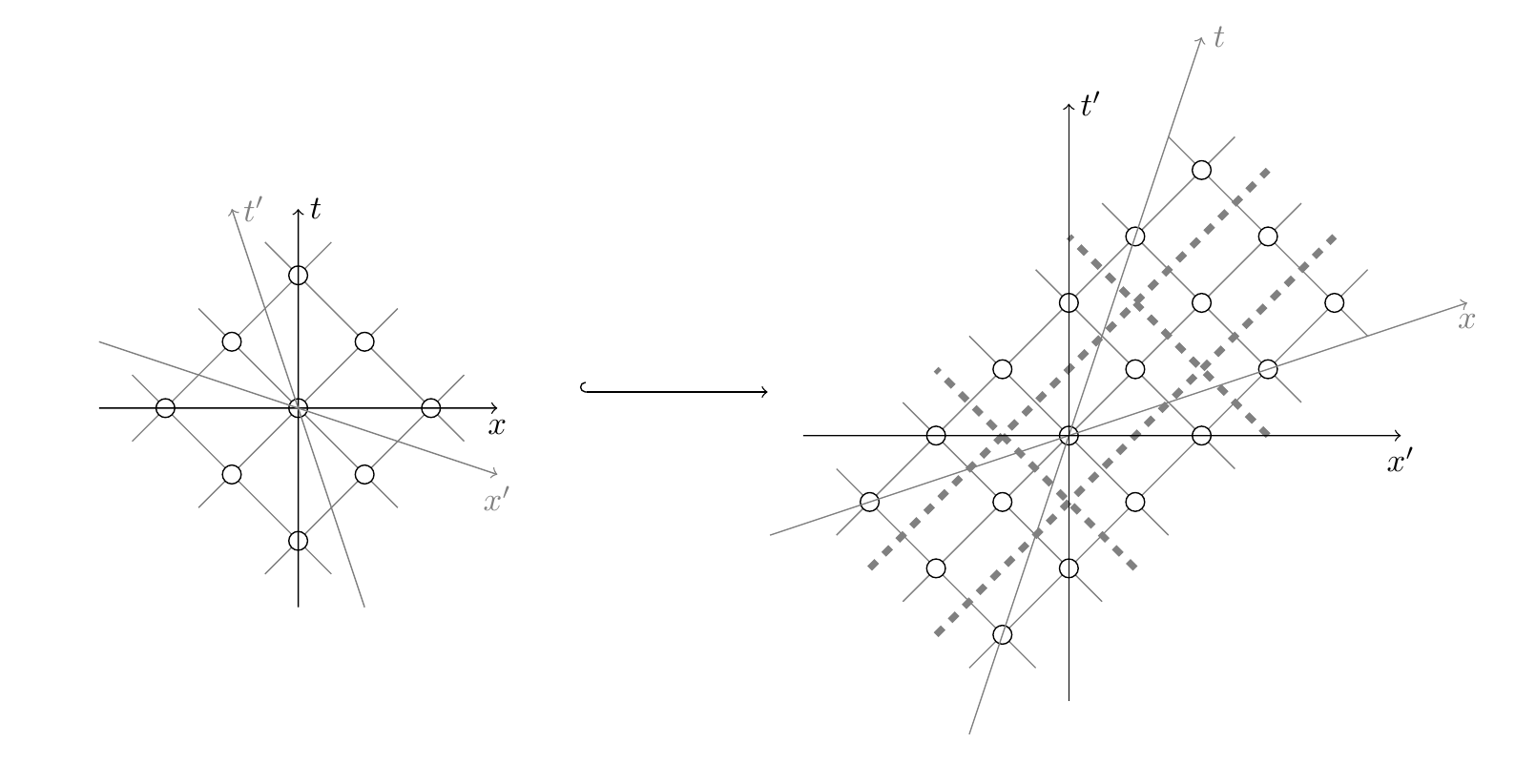}
\caption{\label{fig:boostedaxes} {\em Conceptual diagram for the discrete Lorentz transform.} In this example $\alpha=2$, $\beta=1$. Each point in the original reference frame is transformed into a lightlike rectangular spacetime patch of $\alpha\times\beta$ points, here enclosed by the dashed lines. This switches from the $x,t$ frame to the $x',t'$ frame, as shown.}
\end{figure*}

By definition, therefore, they have several of the fundamental symmetries  of physics, built-in. But can they also have Lorentz covariance? The purpose of this article is to address this question. 

{\em Summary of results.}
Lorentz covariance states that the laws of physics remain the same in all inertial frames. Lorentz transforms relate spacetimes as seen by different inertial frames. This paper formalizes a notion of discrete Lorentz transforms, acting upon wavefunctions over discrete spacetime. It formalizes the notion of discrete Lorentz covariance of a QW, by demanding that a solution of the QW be Lorentz transformed into another solution, of the same QW.

Before the formalism is introduced, the paper investigates a concrete example: the Dirac QW \cite{BenziSucci,Bialynicki-Birula,MeyerQLGI,ArrighiDirac}. The Dirac QW is a natural candidate for being Lorentz covariant, because its continuum limit is the covariant, free-particle Dirac equation \cite{StrauchPhenomena,ArrighiDirac,bisio2013dirac}. This example helps us build our definitions. However, the Dirac QW turns out to be first-order covariant only. In order to obtain exact Lorentz covariance, we introduce a new model, the Clock QW, which arises as the quantum version of a covariant classical Random Walk \cite{wall1988discrete}. However, the Clock QW requires an observer-dependent dimension for the internal state space. In order to overcome this problem, the formalism is extended to multiple-walkers QWs, i.e. Quantum Cellular Automata (QCA). Indeed, the Clock QCA provides a first finite-dimensional model of an exactly covariant QCA. We use numerous figures to help our intuition. In fact, the theory admits a simple diagrammatic representation, in terms of a few local, circuit equivalence rules. The theory also allows for non-homogeneous Lorentz transforms, a specific class of general coordinate transformations, and yet expressive enough to switch between non-inertial frames.

{\em Related works.} Researchers have tried to reconcile discreteness and Lorentz covariance in several ways. 

In the causal set approach, only the causal relations between the spacetime events is given. Without a background spacetime Lorentz covariance is vacuous. If, however, the events are generated from a Poissonian distribution over a flat spacetime, then covariance is recovered in a statistical sense \cite{dowker2004quantum}.

Researchers working on Lattice Boltzmann methods for relativistic hydrodynamics also take a statistical approach: the underlying model breaks Lorentz covariance, but the statistical distributions generated are covariant \cite{mendoza2010derivation}. 

Loop Quantum Gravity offers a deep justification for the statistical approach. By interpreting spacetime intervals as the outcome of measurements of quantum mechanical operators, one can obtain covariance for the mean values, while keeping to a discrete spectrum \cite{rovelli2003reconcile,livine2004lorentz}.

The idea of interpreting space and time as operators with a Lorentz invariant discrete spectrum goes back to Snyder \cite{snyder1947quantized}. This line of research goes under the name of Doubly Special Relativity (DSR). Relations between DSR and QWs are discussed in \cite{bibeau2013doubly}. In the DSR approach, a deformation of the translational sector of the Poincar\'e algebra is required.

Instead of deforming the translation operator algebra, one could look at dropping translational invariance of the QW evolution. Along these lines, models have been constructed for QWs in external fields, including specific cases of gravitational fields \cite{MolfettaDebbasch2014Curved, MolfettaDebbasch}.

Another non-statistical, early approach is to restrict the class of allowed Lorentz transforms, to a subgroup of the Lorentz group whose matrices are over the integers numbers \cite{schild1948discrete}. Unluckily, there are no non-trivial integral Lorentz transforms in (1+1)-dimensions. Moreover, interaction rules that are covariant under this subgroup are difficult to find \cite{tHoofttwodimensional,das1960cellular}.

{\em Approach.} 
The approach of the present paper is non-statistical: we look for exact Lorentz covariance. Spacetime remains undeformed, always assumed to be a regular lattice, and the QW remains homogeneous. While keeping to $1+1$ dimensions and integral transforms, we allow for a global rescaling, so that we can represent all Lorentz transforms with rational velocity. The basic idea is to map each point of the lattice to a lightlike rectangular spacetime patch, as illustrated in Fig. \ref{fig:boostedaxes} and \ref{fig:transformingall}.

{\em Plan of the paper.} 
The remainder of this paper is organized as follows. In Section \ref{sec:Preliminaries} we set notations by recalling the Dirac QW and the proof of covariance of the Dirac Eq. In Section \ref{sec:aLorentz transform} we discuss the first-order-only covariance of the Dirac QW. In Section \ref{sec:Lorentztransforms} we formalize discrete Lorentz transforms, covariance, and discuss non-homogeneous Lorentz transforms. In Sections \ref{sec:ClockQW} and \ref{sec:ClockQCA} apply this theory to the Clock QW and the Clock QCA respectively. We finish with a discussion in Section \ref{sec:Conclusions}. 

\section{Preliminaries} \label{sec:Preliminaries}

\subsection{Finite Difference Dirac Eq. and the Dirac QW}\label{subsec:DiracQW}
{\em The Dirac Equation.} 
The (1+1)-dimensional free particle Dirac equation is (with Planck's constant and the velocity of light set to one):
\begin{align}
\partial_t \psi &= -\ii m\sigma_1{\psi} -  \partial_x \sigma_3 \psi, \label{eq:Dirac1D}
\end{align}
where $m$ is the mass of the particle, $\psi=\psi(t,x)$ is a spacetime wavefunction from $\mathbb{\R}^{1+1}$ to $\mathbb{C}^2$ and $\sigma_j$ ($j=0,\ldots,3$) are the Pauli spin matrices, with $\sigma_0$ the identity. Eq. (\ref{eq:Dirac1D}) corresponds to the Weyl (or spinor) representation \cite{Thaller}. 

\medskip  {\em Lightlike coordinates.} 
In order to study covariance, it is always a good idea to switch to lightlike coordinates $r=(t+x)/2$ and $l=(t-x)/2$, in which a Lorentz transform is just a rescaling of the coordinates. Redefine the wavefunction via $\psi(r+l,r-l)\rightarrow \psi(r,l)$, then Eq. \eqref{eq:Dirac1D} becomes
\begin{align}
\diag{\partial_r}{\partial_l} {\psi} &= -\ii m\sigma_1{\psi}. \label{eq:LightDirac1D}
\end{align}

\begin{figure*}[tpb]
  \subfloat[Each white dot (left) represents the corresponding piece of circuit (right).]{\includegraphics[scale=.8]{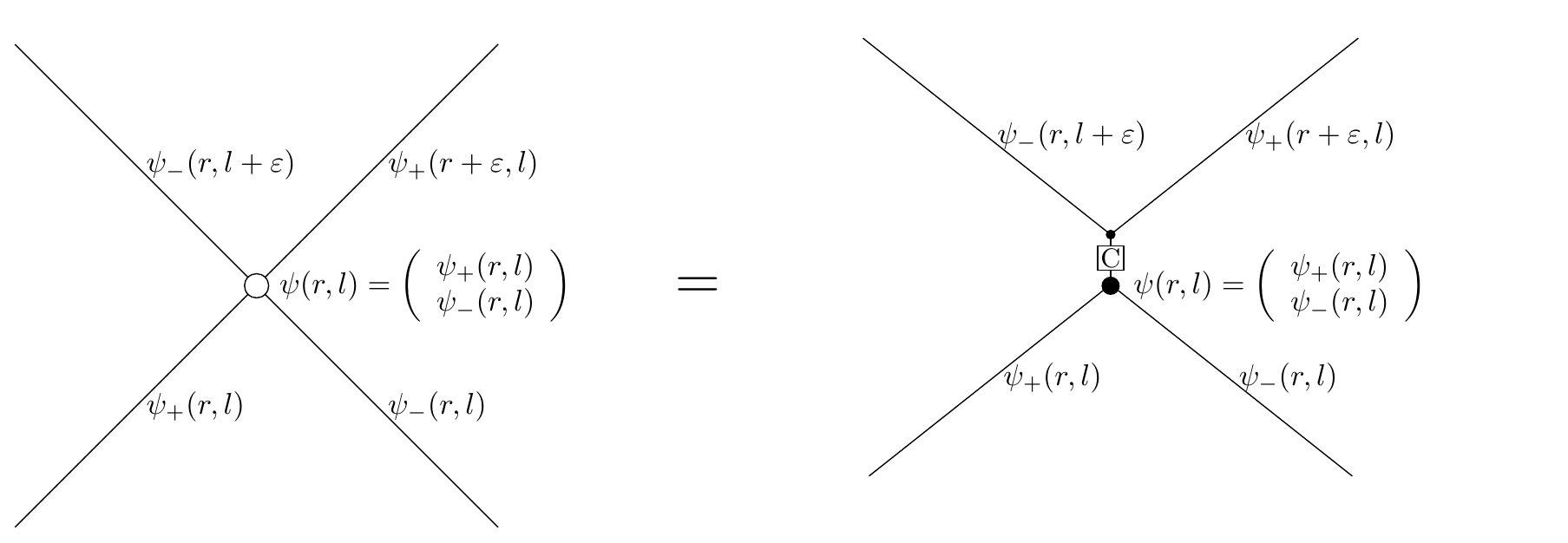}}\quad
  \subfloat[A discrete spacetime wavefunction $\psi$ in lightlike coordinates. Time flows upwards.]{\includegraphics[width=\columnwidth]{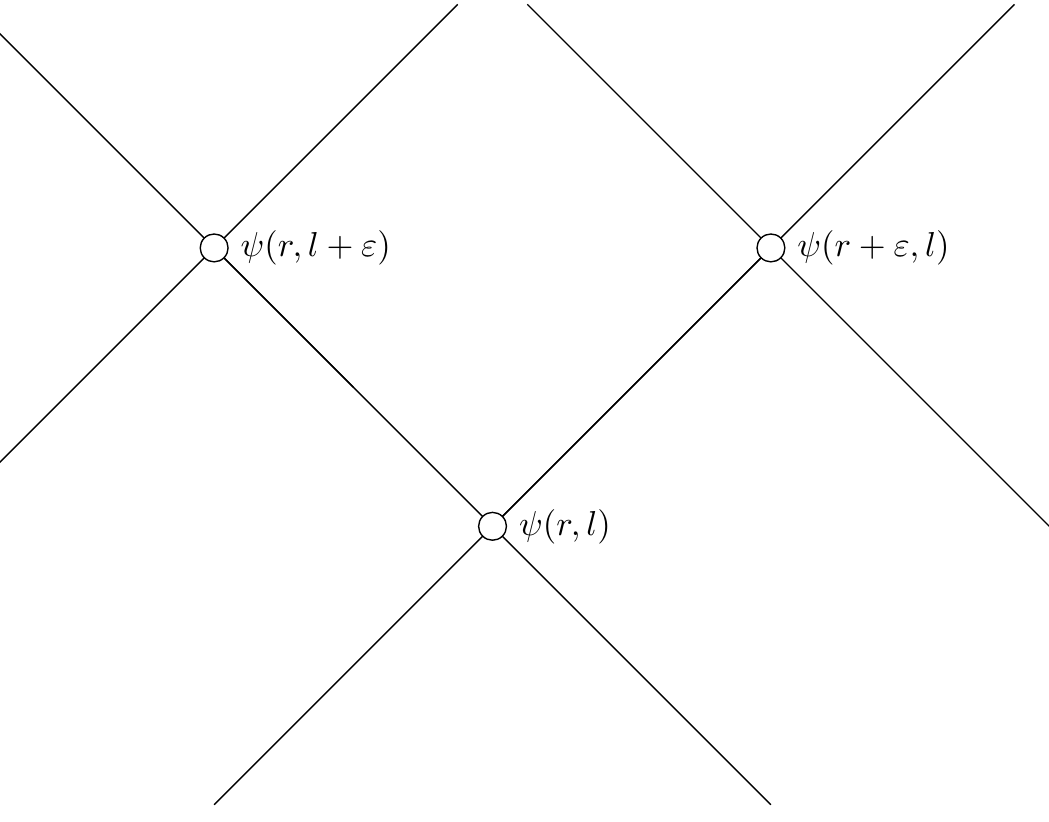}}\quad 
  \subfloat[An explicit circuit-like representation of the relationship between the vectors at each point. The matrix $C$ gets applied upon its input vector, whereas each lightlike wire propagates just one scalar component of the output vector.]{\includegraphics[width=\columnwidth]{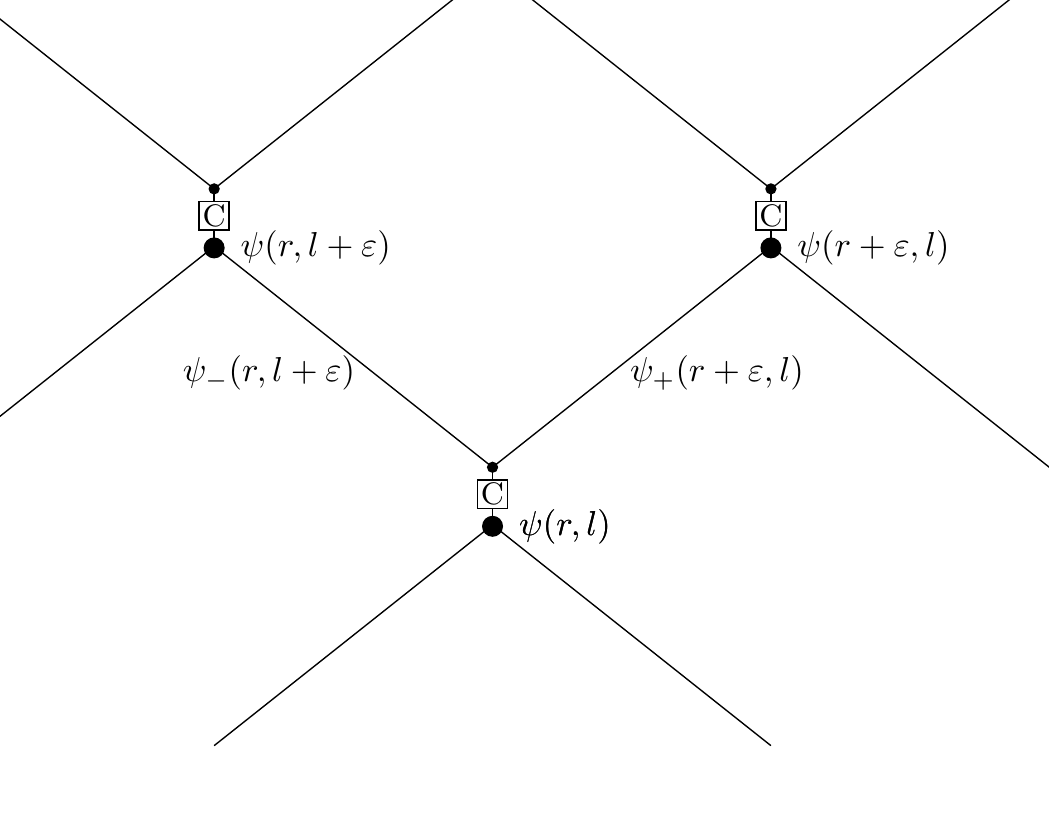}} 
  \caption{\label{fig:conventions} {\em Discrete spacetime wavefunctions that are solutions of the FD Dirac or the Dirac QW.}}
\end{figure*}
 
\medskip {\em Finite-difference Dirac Equation.} 
In this paper $e^{\varepsilon\partial_\mu}$ will be used as a notation for the translation by $\varepsilon$ along the $\mu$-axis (with $\mu=0,1$), i.e. $(e^{\varepsilon\partial_\mu}{\psi})(x_\mu) = {\psi(x_\mu+\varepsilon)}$. 

Using the first order expansion of the the exponential, the spacetime wavefunction $\psi$ is a solution of the Dirac equation if and only if, as $\varepsilon \rightarrow 0$,
\begin{align}\diag{e^{\varepsilon\partial_r}}{e^{\varepsilon\partial_l}}{\psi}
&=\left(\Id+\diag{\varepsilon\partial_r}{\varepsilon\partial_l}\right){\psi}+O(\varepsilon^2)  \nonumber \\
&=(\Id - \ii m \varepsilon \sigma_1){\psi}+O(\varepsilon^2).\label{eq:FDApprox}
\end{align}

Equivalently, if we denote $\psi=(\psi_+,\psi_-)^\mathsf{T}$, then $\psi$ is a solution of the Dirac equation if and only if, to first order in $\varepsilon$ and as $\varepsilon \rightarrow 0$,
\begin{align}
&\psi_+(r+\varepsilon,l)=\psi_+(r,l)-\ii m  \varepsilon \psi_-(r,l) \nonumber \\
\textrm{and}\quad &\psi_-(r,l+\varepsilon)=\psi_-(r,l)-\ii m  \varepsilon \psi_+(r,l). \label{eq:FDDirac1D}
\end{align}
If we now suppose that $\varepsilon$ is fixed, and consider that ${\psi}$ is a spacetime wavefunction from $(\varepsilon\mathbb{Z})^{2}$ to $\mathbb{C}^2$, then Eq. \eqref{eq:FDDirac1D} defines a Finite-difference scheme for the Dirac equation (FD Dirac). As a dynamical system, this FD Dirac is illustrated in Fig. \ref{fig:conventions} with:
\begin{equation}
C=\left(\begin{array}{cc}
1 & -\ii\varepsilon m\\
-\ii\varepsilon m & 1
\end{array} \right). \label{eq:FDQWCoin}
\end{equation}

\medskip {\em The Dirac QW.} 
We could have gone a little further with Eq. (\ref{eq:FDApprox}). Indeed, by  recognizing in the right-hand side of the equation the first order expansion of an exponential, we get:
\begin{align}
\diag{e^{\varepsilon\partial_r}}{e^{\varepsilon\partial_l}}{\psi} &= e^{-\ii m \varepsilon \sigma_1}{\psi} +O(\varepsilon^2). \label{eq:DiracQW}
\end{align}

In fact, $\psi$ is a solution of the Dirac equation if and only if, as $\varepsilon\rightarrow 0$, Eq. (\ref{eq:DiracQW}) is satisfied. See \cite{StrauchPhenomena,ArrighiDirac} for a rigorous, quantified proof of convergence.

If we again say that $\varepsilon$ is fixed, and so that ${\psi}$ is a discrete spacetime wavefunction, then Eq. \eqref{eq:DiracQW} defines a Quantum Walk for the Dirac equation (Dirac QW) \cite{BenziSucci,Bialynicki-Birula,MeyerQLGI,bisio2013dirac,ArrighiDirac}. Indeed, as a dynamical system, this Dirac QW is again illustrated in Fig. \ref{fig:conventions} but this time taking:
\begin{equation}
C=e^{-\ii m \varepsilon \sigma_1} = \left(\begin{array}{cc}
\cos(\varepsilon m) & -\ii\sin(\varepsilon m)\\
-\ii\sin(\varepsilon m) & \cos(\varepsilon m)
\end{array} \right),  \label{eq:DiracQWCoin} 
\end{equation}
which is exactly unitary, i.e. to all orders in $\varepsilon$. 

In the original $(t,x)$ coordinates, both the FD Dirac and the Dirac QW evolutions are given by $\psi(t+\varepsilon,x) = TC\psi(t,x)$, where $T=e^{-\varepsilon \partial_x \sigma_3}$ is the shift operator and $C$ is the matrix appearing in Eq. (\ref{eq:FDQWCoin}) or Eq. (\ref{eq:DiracQWCoin}) respectively (see \cite{ArrighiDirac} for details). In the case of the Dirac QW, $W=TC$ is referred to as the walk operator: it is shift-invariant and unitary. $C$ is referred to as the coin operator, acting over the `coin space', which is ${\cal H} \cong \C^2$ for the Dirac QW. Eq. (\ref{eq:DiracQW}) reads as follows: the top and bottom components of the coin space get mixed up by the coin operator, and then the top component moves at lightspeed towards the right, whereas the bottom component goes in the opposite direction.  

\begin{figure}
\centering
\includegraphics[width=\columnwidth]{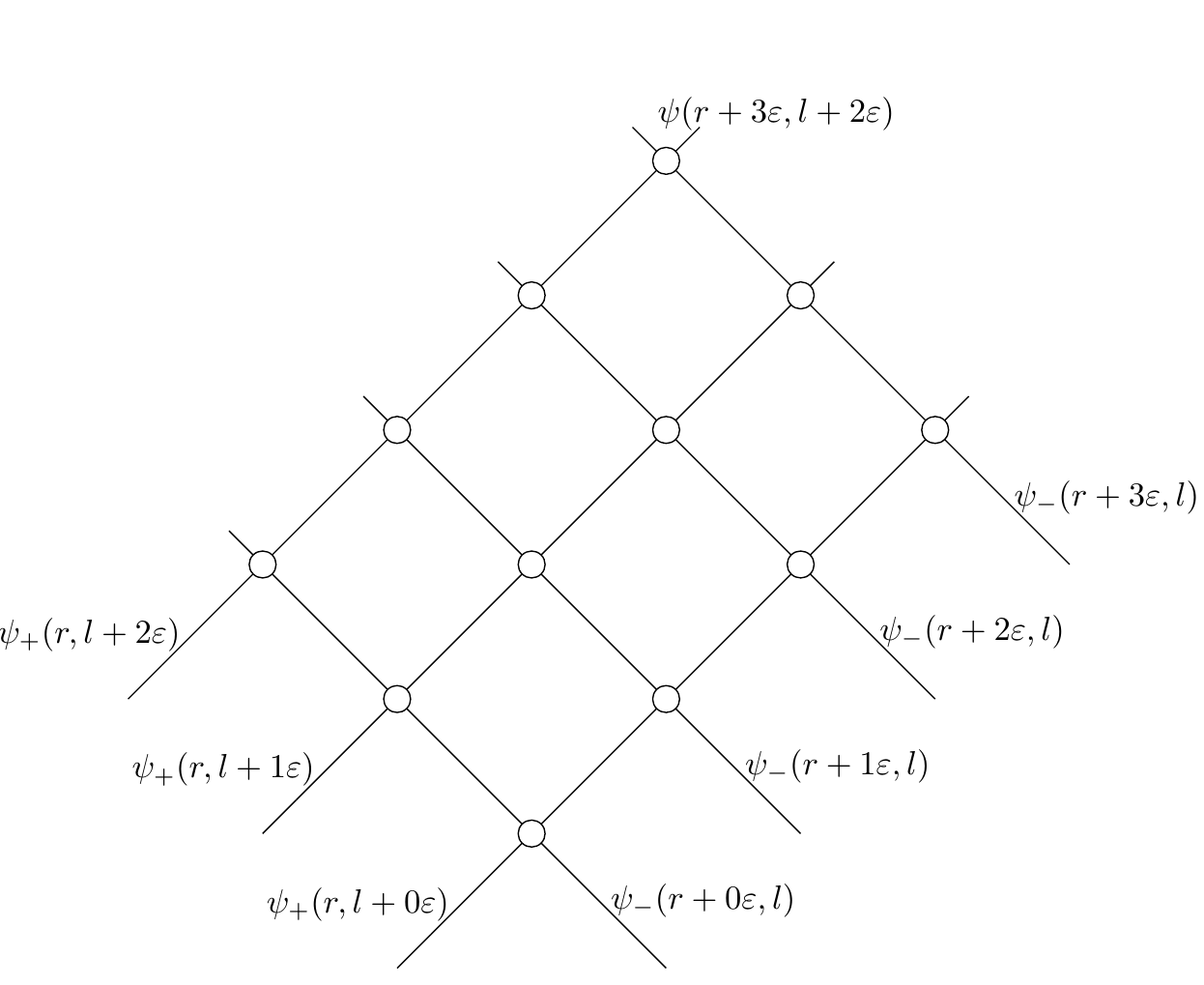}
\caption{\label{fig:determined} {\em Lightlike rectangular patches of spacetime} (in this example $\alpha=4$, $\beta=3$) are fully-determined by the incoming wires.}
\end{figure}

\begin{rk}\label{rk:detregion}  Let $\alpha, \beta$ be arbitrary positive integers. Notice that knowing the value of the scalars $\psi_-(r,l),\ldots,\psi_-(r+(\alpha-1)\varepsilon,l)$ carried by the right-incoming wires, together with the scalars $\psi_+(r,l),\ldots,\psi_+(r,l+(\beta-1)\varepsilon)$ carried by the left-incoming wires, fully determines $\psi(r+i\varepsilon,l+j\varepsilon)$ for $0\leq i \leq (\alpha-1)$ and $0\leq j \leq (\beta-1)$, as made apparent in Fig. \ref{fig:determined}. We denote by $\overline{C}(i,j)$ the operator which, given the vectors
\begin{align*}
\overline{\psi}_-(r,l)=
\left(\begin{array}{c} \psi_-(r,l)\\\vdots\\\psi_-(r+(\alpha-1)\varepsilon,l)\end{array}\right)
\end{align*} 
and
\begin{align*}
\overline{\psi}_+(r,l)=
\left(\begin{array}{c} \psi_+(r,l)\\\vdots\\\psi_+(r,l+(\beta-1)\varepsilon)\end{array}\right)
\end{align*}
combined as
\begin{align*}
\overline{\psi}(r,l)=
\left(\begin{array}{c} \overline{\psi}_+(r,l)\\ \overline{\psi}_-(r,l)\end{array}\right),
\end{align*}
yields $\psi(r+i\varepsilon,l+j\varepsilon)$, i.e. $\psi(r+i\varepsilon,l+j\varepsilon)=\overline{C}(i,j)\overline{\psi}(r,l).$ 
Moreover, notice that those values also determine the right outcoming wires  $\psi_+(r+\alpha\varepsilon,l+j\varepsilon)$ for $0\leq j \leq (\beta-1)$, which we denote by $\overline{C}_+\overline{\psi}(r,l)$, and the left outcoming wires $\psi_-(r+i\varepsilon,l+\beta\varepsilon)$ for $0\leq i \leq (\alpha-1)$, which we denote be $\overline{C}_-\overline{\psi}(r,l)$. More generally, we denote by $\overline{C}$ the circuit made of $(\alpha\beta)$ gates shown in Fig. \ref{fig:determined}, i.e. $$\overline{C}~\overline{\psi}(r,l)=\left(\overline{C}_+ \oplus \overline{C}_-\right)\overline{\psi}(r,l).$$ We write $\overline{C}_m$ for the operator, instead of $\overline{C}$, when we want to make explicit its dependency upon the parameter $m$.
\end{rk}

\subsection{Scaled Lorentz transforms and covariance}\label{subsec:conttransform}

Let us review the covariance of the Dirac equation in a simple manner, that will be useful for us later. Consider a change of coordinates $r'=\alpha r$, $l'=\beta l$. This transformation is proportional by a factor of $\sqrt{\alpha\beta}$ to the Lorentz transform
\begin{equation*}
\Lambda = \left(\begin{array}{cc}
    \sqrt{\dfrac\alpha\beta} & 0 \\
    0 & \sqrt{\dfrac\beta\alpha}
  \end{array}\right)
\end{equation*} 
whose velocity parameter is $u=(\alpha-\beta)/(\alpha+\beta)$. Let us define $\widetilde{\psi}(r',l')=\widetilde{\psi}(\alpha r,\beta l)=\psi(r,l)$. A translation by $\varepsilon$ along $r$ (resp. $l$) becomes a translation by $\alpha \varepsilon$ along $r'$ (resp. $\beta \varepsilon$ along $l'$). Hence the Dirac equation now demands that as $\varepsilon\rightarrow 0$,
\begin{align*}
{\diag{e^{\alpha \varepsilon\partial_{r'}}}{e^{\beta \varepsilon\partial_{l'}}}}{\widetilde{\psi}}
&=
\left(\begin{array}{cc}
1 & -\ii\varepsilon m\\
-\ii\varepsilon m & 1
\end{array} \right)
{\widetilde{\psi}}+O(\varepsilon^2).
\end{align*}
Equivalently, to first order in $\varepsilon$ and as $\varepsilon \rightarrow 0$,
\begin{align*}
\widetilde{\psi}_+(r'+\alpha \varepsilon,l')&=\widetilde{\psi}_+(r',l')-\ii m  \varepsilon \widetilde{\psi}_-(r',l')\\
\widetilde{\psi}_-(r',l'+\beta \varepsilon)&=\widetilde{\psi}_-(r',l')-\ii m  \varepsilon \widetilde{\psi}_+(r',l')
\end{align*}
Unfortunately, whenever $\alpha\neq\beta$, this is not in the form of a Dirac equation. In other words the coordinate change alone does not take the Dirac equation into the Dirac equation. 

\begin{rk} In Section \ref{sec:ClockQW} we will study the Clock QW, inspired by:
\begin{align}
\quad \left(\begin{array}{cc}
e^{\alpha\varepsilon\partial_r} & 0\\
0 & e^{\beta\varepsilon\partial_l}
\end{array} \right) \psi
=e^{-\ii m \varepsilon \sigma_1} \psi.
\end{align}
\end{rk}

Meanwhile, notice that in the first order, the top and bottom $\varepsilon$ can be taken to be different, leading to

\begin{align*}
&{\diag{e^{\varepsilon\partial_{r'}}}{e^{\varepsilon\partial_{l'}}}}{\widetilde{\psi}}
=
\left(\begin{array}{cc}
1 & -\ii \varepsilon m/\alpha\\
-\ii\varepsilon m/\beta & 1
\end{array} \right)
{\widetilde{\psi}}+O(\varepsilon^2).
\end{align*}
\begin{align*}
&{\diag{e^{\varepsilon\partial_{r'}}}{e^{\varepsilon\partial_{l'}}}}
\left(\begin{array}{c}
\widetilde{\psi}_+ / \sqrt{\beta}\\
\widetilde{\psi}_- / \sqrt{\alpha}
\end{array} \right)= \\
&\left(\begin{array}{cc}
1 & -\ii \varepsilon m/\sqrt{\alpha\beta}\\
-\ii\varepsilon m/\sqrt{\alpha\beta} & 1
\end{array} \right)
\left(\begin{array}{c}
\widetilde{\psi}_+/\sqrt{\beta}\\
\widetilde{\psi}_-/\sqrt{\alpha}
\end{array} \right)+O(\varepsilon^2).
\end{align*}

Let us define 
\begin{align*}
S=\left(\begin{array}{cc}
1/\sqrt{\beta} & 0\\
0 & 1/\sqrt{\alpha}
\end{array} \right)\qquad\textrm{and}\qquad \psi'=S\widetilde{\psi}.
\end{align*}
Call this $\psi'$ the Lorentz transformed of $\psi$, instead of $\widetilde{\psi}$. Now we have: 
\begin{align*}
&{\diag{e^{\varepsilon\partial_{r'}}}{e^{\varepsilon\partial_{l'}}}}{\psi'}
=
\left(\begin{array}{cc}
1 & -\ii\varepsilon m/\sqrt{\alpha\beta}\\
-\ii\varepsilon m/\sqrt{\alpha\beta} & 1
\end{array} \right)
{\psi'}\\
\end{align*}
i.e. the Dirac equation just for a different mass $m' = m / \sqrt{\alpha \beta}$. This different mass is due to the fact that the transformation to primed coordinates that we considered was a scaled a Lorentz transform. In the special case where $\alpha \beta = 1$, the above is just the proof of Lorentz covariance of the Dirac equation.

\section{A discrete Lorentz transform for the Dirac QW}\label{sec:aLorentz transform}

\begin{figure*}
\includegraphics[scale=.6]{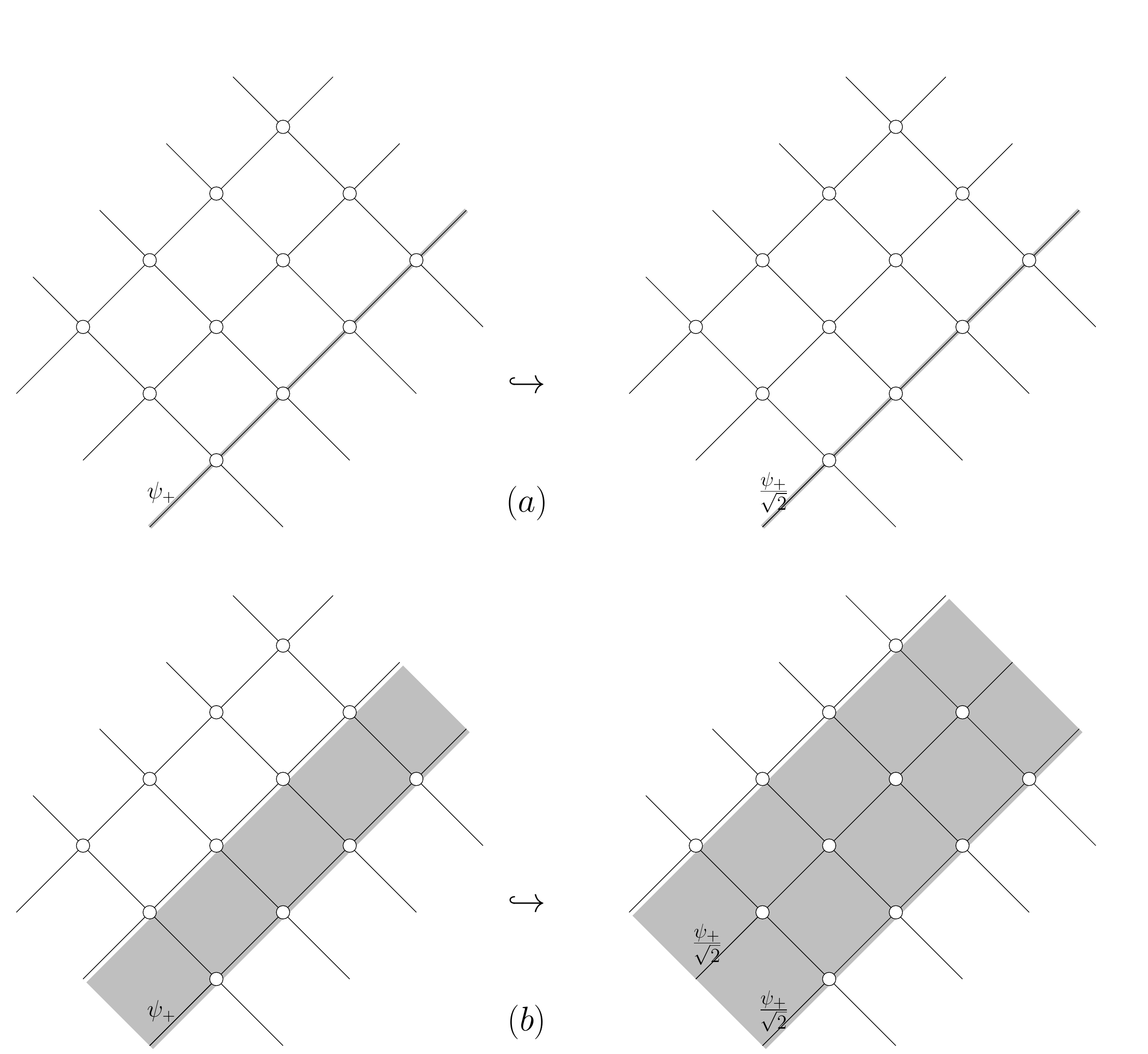}
\caption{\label{fig:normalizationproblem} {\em The normalization problem and solution in the $m=0$, $\alpha=1$, $\beta=2$ case.} $(a)$ If $\psi_+(0,0)$ gets interpreted as a right-traveling Dirac peak, then its transformed version is $\psi_+'(0,0)=\psi_+(0,0)/\sqrt{2}$, which is not normalized. $(b)$ If $\psi_+(0,0)$ gets interpreted as a right-moving rectangular function, then its transformed version spreads out as $\psi_+'(0,0)=\psi_+'(0,1)=\psi_+(0,0)/\sqrt{2}$, which is normalized.}
\end{figure*}

\subsection{Normalization problem and its solution}\label{subsec:normalization}
\medskip {\em Normalization problem in the discrete case.}
Take $\psi(r,l)$ a solution of the Dirac QW such that the initial condition is normalized and localized at single point e.g. $\psi(0,0) = (1,0)^\mathsf{T}$ and $\psi(r,l)=(0,0)^\mathsf{T}$ for $t=r+l=0$. Then, after applying the Lorentz transform described in Subsection \ref{subsec:conttransform}, the initial condition is $\psi'(0,0) = (1/\sqrt\beta,0)^\mathsf{T}$ and $\psi'(r,l)=(0,0)^\mathsf{T}$ for $t=r+l=0$ which is not normalized for any non-trivial Lorentz transform, see Fig. \ref{fig:normalizationproblem}$(a)$. Hence, we see that the Lorentz transform described in Subsection \ref{subsec:conttransform}, i.e. that used for the covariance of the continuous Dirac equation, is problematic in the discrete case: the transformed observer sees a wavefunction which is not normalized. This seems a paradoxical situation since in the limit when $\varepsilon\to 0$, the discrete case tends towards the continuous case, which does not have such a normalization issue. In order to fix this problem, let us look more closely at how normalization is preserved in the continuous case.

\medskip {\em Normalization in the continuous case.}
Now take $\psi(r,l)$ a solution of the massless Dirac equation such that the initial condition is the normalized, right-moving rectangular function, i.e. $\psi(r,l)=(1/\sqrt{2},0)^\mathsf{T}$, for $0 \leq l < 1$ and $\psi(r,l)= (0,0)^\mathsf{T}$ elsewhere. The Lorentz transformed of $\psi$ is
\begin{equation*}
\psi'(r',l') = S\psi(r'/\alpha, l'/\beta) = \left\{ \begin{array}{ll} \left(\begin{array}{c} 1/\sqrt{2\beta}\\0\end{array}\right) & 0 \leq l' < \beta \\ \left(\begin{array}{c} 0\\0\end{array}\right) & \mbox{elsewhere},\end{array}\right.
\end{equation*}
which is normalized. We see that the $S$ matrix is no longer a problem for normalization, but rather it is needed to compensate for the larger spread of the wavefunction, see Fig. \ref{fig:normalizationproblem}$(b)$.
This suggests that the normalization problem for the localized initial condition in the discrete case could be fixed, by allowing the discrete Lorentz transform to spread out the initial condition.

\medskip {\em From the continuous to the discrete.} Intuitively, we could think of defining the discrete Lorentz transform as the missing arrow ``$\textrm{Discrete } \Lambda?$'' that would make the following diagram commute:
\begin{equation*}
\begin{tikzcd}
Dirac \arrow{r}{\Lambda} & Dirac' \arrow{d}{\textrm{Sample}} \\
QW \arrow{r}{\textrm{Discrete }\Lambda?} \arrow{u}{\textrm{Interpolate}} & QW'
\end{tikzcd}.
\end{equation*}
In other words,
\begin{equation*}
\textrm{Discrete }\Lambda:= \textrm{Sample} \circ \Lambda \circ \textrm{Interpolate}
\end{equation*}
The discrete Lorentz transform that we propose next originates from this idea, even though it is phrased directly in the discrete setting. Later in Section \ref{sec:Lorentztransforms} we provide a more general and diagrammatic definition of discrete Lorentz transform and discrete Lorentz covariance.

\subsection{A discrete Lorentz transform} \label{subsec:DiracLorentzTransform}
\begin{figure*}[tbp]
	\subfloat[Individual points and pairs incoming wires of the original spacetime diagram.]{\includegraphics[scale=0.7]{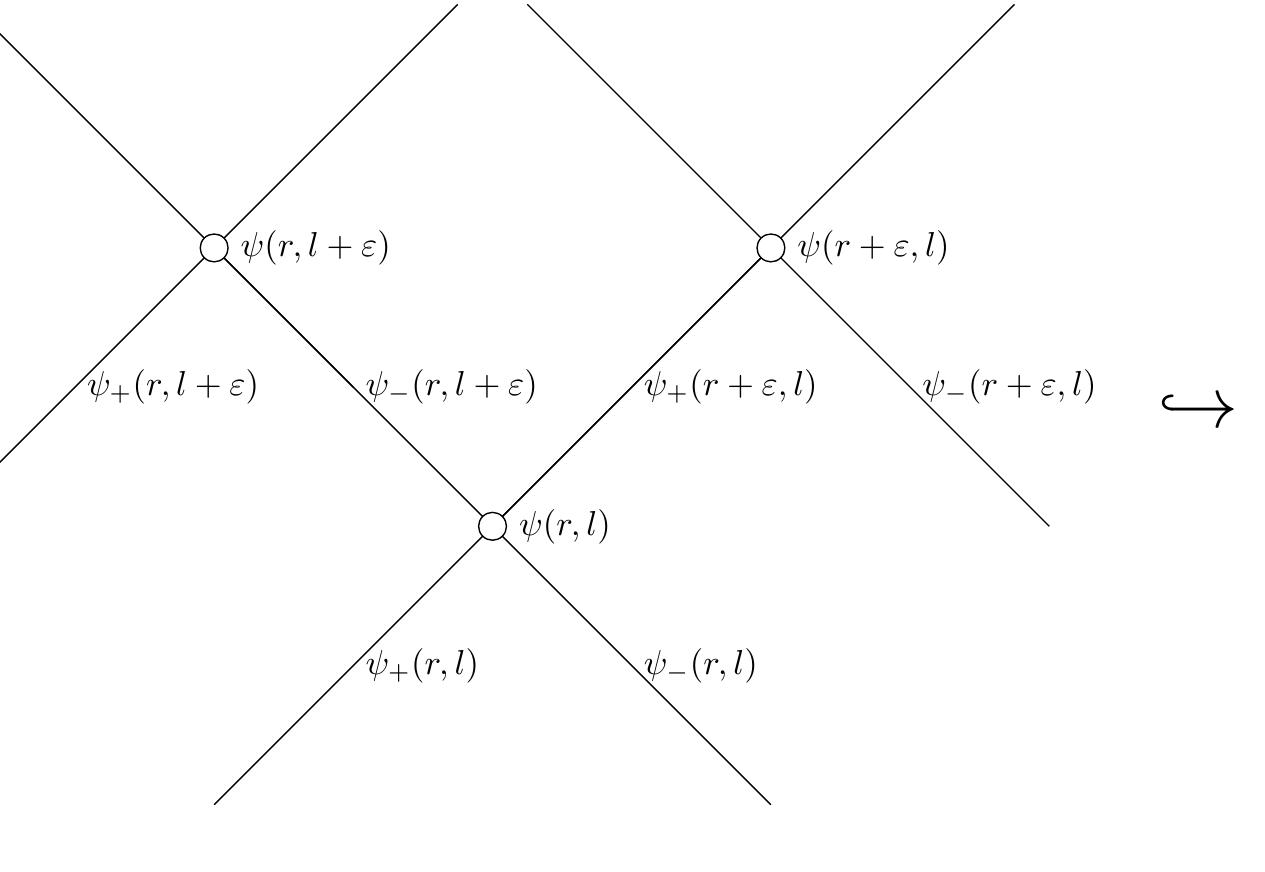}}\quad
	\subfloat[Replacement by a rectangular patch of spacetime, which is a zoomed-in version of the point obtained by spreading out its incoming wires.]{\includegraphics[scale=.8]{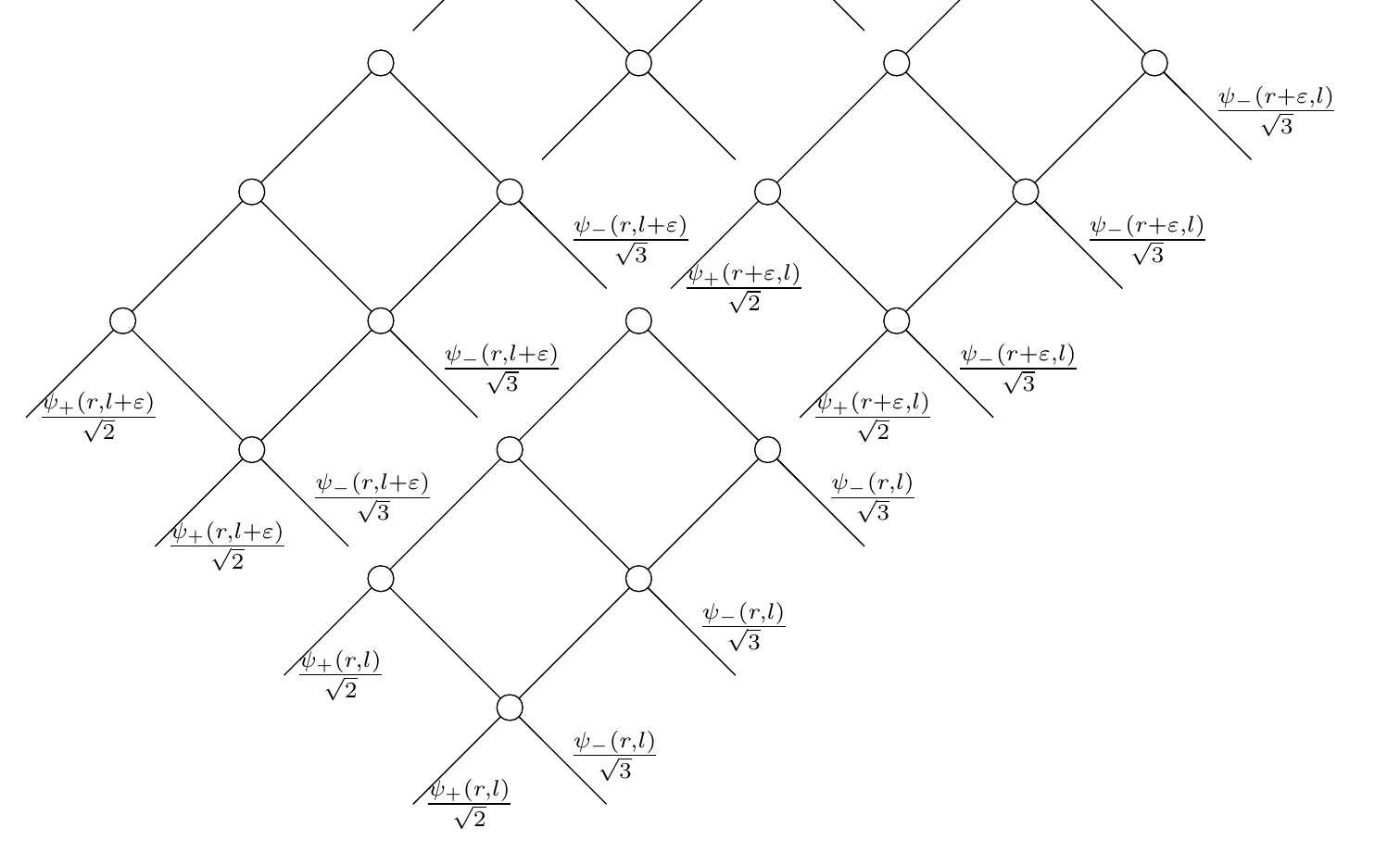}}\\ 
	\caption{\label{fig:Lorentz transform} {\em A discrete Lorentz transform}, with parameters $\alpha=3$, $\beta=2$.}
\end{figure*}

In the continuous case we had $\psi'(r',l')=S\psi(r,l)$. Hence $\psi'(r',l')=S\psi(r'/\alpha,l'/\beta)$. In the discrete case, however ${\psi}$ is a spacetime wavefunction from $(\varepsilon\mathbb{Z})^{2}$ to $\mathbb{C}^2$, as in Fig. \ref{fig:conventions}$(b)$. Hence, demanding, for instance, that $\psi'(\varepsilon,0)=S\psi(\varepsilon/\alpha,0)$ becomes meaningless, because $\psi(\varepsilon/\alpha,0)$ is undefined. The normalization issues and the related discussion of Subsection \ref{subsec:normalization} suggests setting $\psi_-'(\varepsilon,0)$ to $S\psi_-(0,0)$, and not to $0$. More generally, we will take:
\begin{equation*}
\forall r'\in \varepsilon\alpha \mathbb{Z},\quad\psi_+'(r',l')=\frac{\psi_+( r'/\alpha,\lfloor l'/\beta\rfloor_\varepsilon)}{\sqrt{\beta}}
\end{equation*}
and
\begin{equation*}
\forall l'\in \varepsilon\beta \mathbb{Z},
\quad \psi_-'(r',l')=\frac{\psi_-(\lfloor r'/\alpha\rfloor_\varepsilon, l'/\beta)}{\sqrt{\alpha}}.
\end{equation*}
where $\lfloor .\rfloor_\varepsilon$ takes the closest multiple of $\varepsilon$ that is less or equal to the number. Notice that this implies that for all $r'\in \varepsilon\alpha \mathbb{Z}$ and $l'\in \varepsilon\beta \mathbb{Z}$, we have $\psi'(r',l')=S\psi(r'/\alpha,l'/\beta)$, as in the continuous case. However, what if we have neither $r'\in \alpha\varepsilon\mathbb{Z}$ nor $l'\in \beta\varepsilon\mathbb{Z}$? As was illustrated in Fig. \ref{fig:determined}, this spacetime region is now fully determined, i.e. we set 
\begin{align}
\forall r',l'\in \varepsilon\mathbb{Z},\quad \psi'(r',l')=\overline{C}_{m'}(i,j)\overline{\psi'}(\lfloor r'\rfloor_{\alpha\varepsilon},\lfloor l'\rfloor_{\beta\varepsilon}) \label{eq:Lorentz transform}
\end{align}  
with $m' = m / \sqrt{\alpha \beta}$, $i\varepsilon=r'-\lfloor r'\rfloor_{\alpha\varepsilon}$, $j\varepsilon=l'-\lfloor l'\rfloor_{\beta\varepsilon}$, $\overline{C}_{m'}(i,j)$ as defined in Remark \ref{rk:detregion}, and $\overline{\psi'}(\lfloor r'\rfloor_{\alpha\varepsilon},\lfloor l'\rfloor_{\beta\varepsilon})$ again as defined in Remark \ref{rk:detregion}, namely
\begin{align*}
\overline{\psi'}_+(\lfloor r'\rfloor_{\alpha\varepsilon},\lfloor l'\rfloor_{\beta\varepsilon})
&=\left(\begin{array}{c} \psi'_+(\lfloor r'\rfloor_{\alpha\varepsilon},\lfloor l'\rfloor_{\beta\varepsilon})\\\vdots\\ \psi'_+(\lfloor r'\rfloor_{\alpha\varepsilon},\lfloor l'\rfloor_{\beta\varepsilon}+(\beta-1)\varepsilon ) \end{array}\right) \\
&=\left(\begin{array}{c} \dfrac{\psi_+(\lfloor r'\rfloor_{\alpha\varepsilon},\lfloor l'\rfloor_{\beta\varepsilon})}{\sqrt{\beta}}\\\vdots\end{array}\right) 
\end{align*}
and similarly
\begin{align*}
\overline{\psi'}_-(\lfloor r'\rfloor_{\alpha\varepsilon},\lfloor l'\rfloor_{\beta\varepsilon})&=
\left(\begin{array}{c} \dfrac{\psi_-(\lfloor r'\rfloor_{\alpha\varepsilon},\lfloor l'\rfloor_{\beta\varepsilon})}{\sqrt{\alpha}}\\\vdots\end{array}\right) 
\end{align*}
and finally
\begin{align*}
\overline{\psi'}(\lfloor r'\rfloor_{\alpha\varepsilon},\lfloor l'\rfloor_{\beta\varepsilon})=
\left(\begin{array}{c} \overline{\psi'}_+(\lfloor r'\rfloor_{\alpha\varepsilon},\lfloor l'\rfloor_{\beta\varepsilon})\\ \overline{\psi'}_-(\lfloor r'\rfloor_{\alpha\varepsilon},\lfloor l'\rfloor_{\beta\varepsilon})\end{array}\right).
\end{align*} 
This finishes to define a discrete Lorentz transform $L_{\alpha,\beta}$, which is illustrated in Fig. \ref{fig:Lorentz transform}. 

An equivalent, more concise way of specifying this discrete Lorentz transform $L_{\alpha,\beta}$ is as follows. First, consider the isometry $E_\beta$ (resp. $E_\alpha$)  which codes $\psi_+(r,l)$ (resp. $\psi_-(r,l)$) into the more spread out $\widecheck{\psi}_+(r,l)=E_\beta\psi_+(r,l)=\overline{\psi'}_+(\alpha r, \beta l)$ (resp. $\widecheck{\psi}_-(r,l)=E_\alpha\psi_-(r,l)= \overline{\psi'}_-(\alpha r, \beta l)$), and let $\widecheck{\psi}(r,l)=\widecheck{\psi}_+(r,l)\oplus \widecheck{\psi}_-(r,l)$, and $m' = m / \sqrt{\alpha \beta}$. Second, construct $\psi'=L_{\alpha,\beta}\psi$ by replacing every spacetime point $\psi(r,l)$ with the lightlike rectangular spacetime patch $\left(\overline{C}_{m'}(i,j)\widecheck{\psi}(r,l)\right)_{i=0\ldots(\alpha-1),j=0\ldots(\beta-1)}$.

Does this discrete Lorentz transform fix the normalization problem of Subsection \ref{subsec:normalization}? Let us evaluate this question.

\subsection{From continuous to discrete current and norm}\label{subsec:discretecurrent}
\subsubsection{Continuous current and norm}
In order to evaluate the norm of a spacetime wavefunction $\psi$ in the continuous setting, we need the following definition. We say that a surface $\sigma$ is a {\em Cauchy surface} if it intersects every causal curve exactly once (a causal curve being a curve whose tangent vector is always timelike or lightlike). The relativistic current $j^\mu=(j^0,j^1)$ is equal to $j^{\mu}=(|\psi_+|^2 + |\psi_-|^2, |\psi_+|^2 - |\psi_-|^2)$, and in lightlike coordinates becomes $j^{s} = (|\psi_+|^2, |\psi_-|^2)$, $s=\pm$. The norm of $\psi$ along a Cauchy surface $\sigma$ is defined by integrating the current $j^s$ along $\sigma$
\begin{equation}
||\psi||_\sigma^2 = \int_\sigma j^s n_s d\sigma
\end{equation}
where $n_s$ is the unit normal vector to $\sigma$ in $r,l$ coordinates. 

If $\psi$ is a solution of the Dirac equation, then this definition does not actually depend on the surface $\sigma$ (for a proof see for instance \cite{schweber1961introduction}, Chap. 4), and so in this case we can write $||\psi||_\sigma^2=||\psi||^2$.

This definition of norm is Lorentz invariant, indeed:
\begin{align}
||\psi||_{\sigma}^2 &=  \int_\sigma j^s n_s d\sigma \nonumber \\  
&= \int_\sigma \left(\frac{|\psi_+|^2}{\beta}\beta dl + \frac{|\psi_-|^2}{\alpha} \alpha dr\right)  \nonumber  \\
&= \int_{\sigma'} (|\psi'_+|^2dl' + |\psi'_-|^2 dr') \nonumber   \\
&= \int_{\sigma'} j'^s n'_s d\sigma'  \nonumber  \\
&= ||\psi'||_{\sigma'}^2 \label{eq:normlorentzinv}
\end{align}
\subsubsection{Discrete Cauchy surfaces}
\begin{figure}
$(a)$ \begin{tikzpicture}
  \path[draw] (1,1) -- (2,2) -- (3,1);
  \draw[fill=white] (1,1) circle(.08);
  \draw[fill=white] (2,2) circle(.08);
  \draw[fill=white] (3,1) circle(.08);
  \path[draw,color=gray] (1,2) -- (2,1) -- (3,2);
  \node[xshift=30*\gs,yshift=33*\gs]{$(r_0,l_0)$};
  \node[xshift=55*\gs,yshift=27*\gs,color=gray]{$\sigma(0)=L$};
  \node[xshift=\gs,yshift=27*\gs,color=gray]{$\sigma(-1)=R$};
\end{tikzpicture}
\qquad \qquad \newline
$(b)$ \begin{tikzpicture}
  \path[draw] (0,2) -- (2,4);
  \path[draw] (1,1) -- (4,4);

  \path[draw] (0,4) -- (3,1);
  \path[draw] (2,4) -- (4,2);
  \path[draw] (4,4) -- (5,3);

  \draw[fill=white] (0,2) circle(.08);
  \draw[fill=white] (0,4) circle(.08);
  \draw[fill=white] (1,1) circle(.08);
  \draw[fill=white] (1,3) circle(.08);
  \draw[fill=white] (2,2) circle(.08);
  \draw[fill=white] (2,4) circle(.08);
  \draw[fill=white] (3,1) circle(.08);
  \draw[fill=white] (3,3) circle(.08);
  \draw[fill=white] (4,2) circle(.08);
  \draw[fill=white] (4,4) circle(.08);
  \draw[fill=white] (5,3) circle(.08);

  \path[draw,color=gray] (0,3) -- (2,1) -- (5,4);
  \draw[color=gray,fill=gray] (1,2) circle(.05);
  \draw[color=gray,fill=gray] (2,1) circle(.05);
  \draw[color=gray,fill=gray] (3,2) circle(.05);
  \draw[color=gray,fill=gray] (4,3) circle(.05);

  \node[xshift=30*\gs,yshift=33*\gs]{$(r'_0,l'_0)$};
  \node[xshift=50*\gs,yshift=25*\gs,color=gray]{$\alpha=3$};
  \node[xshift=8*\gs,yshift=25*\gs,color=gray]{$\beta=2$};
\end{tikzpicture}
\caption{\label{fig:cauchysurface} {\em Discrete Cauchy surfaces and their transformations.} {\em $(a)$} Centering on the origin $(r_0,l_0)$ of the discrete Cauchy surface. {\em $(b)$} Lorentz transform of the same piece of surface for $\alpha=3,\beta=2$.}
\end{figure}
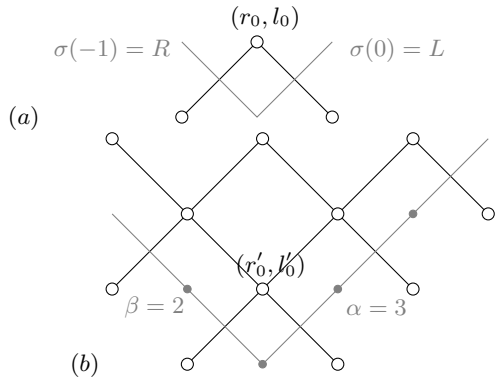

We now provide discrete counterparts to the above notions, beginning with {\em discrete Cauchy surfaces}. Let us consider a function $\sigma: \mathbb{Z} \to \{R,L\}$, and an origin $(r_0,l_0)$. Together, they describe a piecewise linear curve made up of segments of the following form (in gray):\\
\[
\begin{tikzpicture}
  \path[draw] (1,1) -- (2,2);
  \draw[fill=white] (1,1) circle(.08);
  \draw[fill=white] (2,2) circle(.08);
  \path[draw,color=gray] (1,2) -- (2,1);
  \node[xshift=31*\gs,yshift=15*\gs,color=gray]{$R$};
\end{tikzpicture}
\quad
\begin{tikzpicture}
  \path[draw] (2,1) -- (1,2);
  \draw[fill=white] (2,1) circle(.08);
  \draw[fill=white] (1,2) circle(.08);
  \path[draw,color=gray] (1,1) -- (2,2);
  \node[xshift=11*\gs,yshift=15*\gs,color=gray]{$L$};
\end{tikzpicture}
\]
i.e. this curve intersects the spacetime lattice in two ways, labeled $R$ and $L$ (right, left). The centering on the origin is done as in Fig. \ref{fig:cauchysurface} (a).
We say that such a curve is a {\em discrete Cauchy surface} if it does not contain infinite sequences of contiguous $R$ or $L$. One can easily see that such a surface must intersect every lightlike curve exactly once. For concreteness, notice that the discrete equivalent to the continuous constant-time $t=0$ Cauchy surface, is described by:
\begin{equation*}
  \sigma(n) = \left\{
    \begin{array}{ll}
      L & \quad \mbox{for even } n \\
      R & \quad \mbox{for odd } n
    \end{array}
  \right.
\end{equation*}
with origin $(0,0)$.

\subsubsection{Discrete current and discrete norm}
Similarly, let us define the discrete current carried by a wavefunction $\psi$. At each wire connecting two points of the discrete lattice, the current is given by:
\[
\begin{tikzpicture}
  \path[draw] (2*\gs,\gs) -- (\gs,2*\gs);
  \node[xshift=72*\gs,yshift=45*\gs]{$j=|\psi_-(r,l+\varepsilon)|^2$};
  \draw[fill=white] (2*\gs,\gs) circle(.08);
  \draw[fill=white] (\gs,2*\gs) circle(.08);
  \node[xshift=20*\gs,yshift=60*\gs]{$(r,l+\varepsilon)$};
  \node[xshift=65*\gs,yshift=25*\gs]{$(r,l)$};
\end{tikzpicture}
\]

\[
\begin{tikzpicture}
  \path[draw] (\gs,\gs) -- (2*\gs,2*\gs);
  \node[xshift=75*\gs,yshift=45*\gs]{$j=|\psi_+(r+\varepsilon,l)|^2$};
  \draw[fill=white] (\gs,\gs) circle(.08);
  \draw[fill=white] (2*\gs,2*\gs) circle(.08);
  \node[xshift=65*\gs,yshift=60*\gs]{$(r+\varepsilon,l)$};
  \node[xshift=20*\gs,yshift=25*\gs]{$(r,l)$};
\end{tikzpicture}
\]
In analogy with the continuous case, we can evaluate the norm of $\psi$ along a surface $\sigma$ as follows
\begin{equation}
||\psi||_\sigma^2 = \sum_{i\in \mathbb{Z}} j(i)
\end{equation}
where $j(i)$ is the current of the wire at intersection $i$.
For instance, for the discrete constant-time surface the above expression evaluates to the usual $L_2$-norm of a spacelike wavefunction
\begin{align*}
||\psi||_{t=0}^2 &= \sum_{i\in \mathbb{Z}} j(i)\\
&= \sum_{i\in 2\mathbb{Z}+1}|\psi_+(i,-i)|^2 + \sum_{i\in 2\mathbb{Z}}|\psi_-(i,-i)|^2 \\
&= \sum_i ||\psi(i,-i)||^2=||\psi||^2.
\end{align*}

\subsubsection{Cauchy surface independence of the discrete norm}
If $\psi$ is a solution of a QW, then just like the continuous case, the discrete norm does not depend on the discrete Cauchy surface chosen for evaluating it. The proof outline is as follows. First, two Cauchy surfaces $\sigma$ and $\sigma'$ can be made to coincide on an arbitrary large region using only a finite sequence of swap moves\footnote{We take the convention that if the swap move is applied to a segment around the origin, the origin moves along.}:

\[
\begin{array}{c}

\begin{array}{c}
\begin{tikzpicture}
  \path[draw] (1*\gs,1*\gs) -- (2*\gs,2*\gs);
  \path[draw] (2*\gs,1*\gs) -- (1*\gs,2*\gs);
  \draw[fill=white] (1*\gs,1*\gs) circle(.08);
  \draw[fill=white] (2*\gs,2*\gs) circle(.08);
  \draw[fill=white] (1*\gs,2*\gs) circle(.08);
  \draw[fill=white] (2*\gs,1*\gs) circle(.08);
  \draw[fill=white] (1.5*\gs,1.5*\gs) circle(.08);
  \path[draw,color=gray] (1*\gs,1.5*\gs) -- (1.5*\gs,2*\gs) -- (2*\gs,1.5*\gs);
  \node[xshift=20*\gs,yshift=60*\gs]{$(r,l+\varepsilon)$};
  \node[xshift=65*\gs,yshift=60*\gs]{$(r+\varepsilon,l)$};
  \node[xshift=50*\gs,yshift=42*\gs]{$(r,l)$};
  \node[xshift=20*\gs,yshift=25*\gs]{$(r-\varepsilon,l)$};
  \node[xshift=65*\gs,yshift=25*\gs]{$(r,l-\varepsilon)$};
\end{tikzpicture}
\end{array} \\ 
\quad \upharpoonleft \downharpoonright \quad \\
\begin{array}{c}
\begin{tikzpicture}
  \path[draw] (\gs,\gs) -- (2*\gs,2*\gs);
  \path[draw] (2*\gs,\gs) -- (\gs,2*\gs);
  \draw[fill=white] (\gs,\gs) circle(.08);
  \draw[fill=white] (2*\gs,2*\gs) circle(.08);
  \draw[fill=white] (\gs,2*\gs) circle(.08);
  \draw[fill=white] (2*\gs,\gs) circle(.08);
  \draw[fill=white] (1.5*\gs,1.5*\gs) circle(.08);
  \path[draw,color=gray] (\gs,1.5*\gs) -- (1.5*\gs,\gs) -- (2*\gs,1.5*\gs);
  \node[xshift=20*\gs,yshift=60*\gs]{$(r,l+\varepsilon)$};
  \node[xshift=65*\gs,yshift=60*\gs]{$(r+\varepsilon,l)$};
  \node[xshift=50*\gs,yshift=42*\gs]{$(r,l)$};
  \node[xshift=20*\gs,yshift=25*\gs]{$(r-\varepsilon,l)$};
  \node[xshift=65*\gs,yshift=25*\gs]{$(r,l-\varepsilon)$};
\end{tikzpicture}
\end{array}.
\end{array}
\]

Second, swap moves leave the norm invariant, because of unitarity of the $C$ gate (see Fig. \ref{fig:conventions}):
\begin{equation*}
|\psi_+(r+\varepsilon,l)|^2 + |\psi_-(r,l+\varepsilon)|^2 = |\psi_+(r,l)|^2 + |\psi_-(r,l)|^2.
\end{equation*}
Third, take a positive $\delta$. By having $\sigma$ to coincide with $\sigma'$ on a large enough region, we obtain that $|(||\psi||_\sigma-||\psi||_{\sigma'})|\leq \delta$. Lastly, since $\delta$ is arbitrary, $||\psi||_\sigma=||\psi||_{\sigma'}$.

\subsubsection{Lorentz invariance of the discrete norm}
Finally, we will prove the analogue of equation \eqref{eq:normlorentzinv} in the discrete setting. First of all we define how a discrete Cauchy surface $\sigma$ transforms under a discrete Lorentz transform with parameters $\alpha, \beta$. The sequence $\sigma'$ is constructed from $\sigma$ by replacing each $L$ by $L^\alpha$ and each $R$ with $R^\beta$, starting from the center. The origin $(r_0,l_0)$ is mapped to the point $(r'_0,l'_0) = (\alpha r_0, \beta l_0)$. For instance, the piece of surface in Fig. \ref{fig:cauchysurface}(a) is transformed as in Fig. \ref{fig:cauchysurface}(b). We obtain (where $\mathcal{R}_\sigma$ and $\mathcal{L}_\sigma$ are the sets of right and left intersections respectively):
\begin{align*}
||\psi||^2 &= \sum j(i) \\
&= \sum_{i\in \mathcal{R}_\sigma}|\psi_+(r_i,l_i)|^2 + \sum_{i\in \mathcal{L}_\sigma}|\psi_-(r_i,l_i)|^2 \\
&= \sum_{i\in \mathcal{R}_\sigma}\beta \left|\frac{\psi_+(r_{i},l_{i})}{\sqrt{\beta}}\right|^2 + \sum_{i\in \mathcal{L}_\sigma}\alpha\left|\frac{\psi_-(r_{i},l_{i})}{\sqrt{\alpha}}\right|^2 \\
&= \sum_{i'\in \mathcal{R}_{\sigma'}}|\psi'_+(r_{i'},l_{i'})|^2 + \sum_{i'\in \mathcal{L}_{\sigma'}}|\psi'_-(r_{i'},l_{i'})|^2 \\
&= \sum j'(i')  = ||\psi'||^2.
\end{align*}

\subsection{The First-order-only Lorentz covariance of the Dirac QW}\label{subsec:focovariance}

In Subsection \ref{subsec:DiracQW} we defined the Dirac QW, and explained when a spacetime wavefunction $\psi$ is a solution for it.
In Subsection \ref{subsec:DiracLorentzTransform} we defined a discrete Lorentz transform, taking a spacetime wavefunction $\psi$ into another spacetime wavefunction $\psi'$. 
In Subsection \ref{subsec:discretecurrent} we showed that this transformation preserves the norm, i.e. $||\psi||_\sigma^2=||\psi'||_{\sigma'}^2$. 
The question that remains is whether the Dirac QW is Lorentz covariant with respect to this discrete Lorentz transform. In other words, is it the case that $\psi'$ is itself a solution of the Dirac QW, for some $m'$? This demand is concrete translation of the main principle of special relativity, stating that the laws of physics (here, the Dirac QW) remain the same in all inertial reference frames (here, those of $\psi$ and $\psi'$).

Recall that the discrete Lorentz transform works by replacing each point of the spacetime lattice by a lightlike rectangular patch of spacetime, which can be understood as a ``biased, zoomed in version" of that point, see Fig. \ref{fig:Lorentz transform}. Internally, each patch is a piece of spacetime solution of the Dirac QW by construction, see Eq. \eqref{eq:Lorentz transform}. But is it the case that the patches match up, to form the entire spacetime wavefunction of a solution? After all, there could be inconsistencies in between patches: values carried by the incoming wires to the next patches, e.g. $\widecheck{\psi}_+(r+\varepsilon, l)$ (resp. $\widecheck{\psi}_-(r,l+\varepsilon)$) could be different from those carried by the wires coming out of the preceding patch, i.e. $\widehat{\psi}_+(r,l)=\overline{C}_+\widecheck{\psi}(r,l)$ (resp. $\widehat{\psi}_-(r,l)=\overline{C}_-\widecheck{\psi}(r,l)$). 
More precisely, we need both $\widehat{\psi}_+(r,l)=\widecheck{\psi}_+(r+\varepsilon,l)$ and $\widehat{\psi}_-(r,l)=\widecheck{\psi}_-(r,l+\varepsilon)$ for every $r,l$.
This potential mismatch is represented by the discontinuations of the wires of Fig. \ref{fig:Lorentz transform}$(b)$. 
Clearly, the patches making up $\psi'$ match up to form the spacetime wavefunction of a solution if and only if there are no such inconsistencies. We now evaluate these inconsistencies. 

In the first order, the Dirac QW and the Finite-Difference Dirac equation are equivalent, as shown in Subsection \ref{subsec:DiracQW}. This makes it easier to compute the outcoming values of the patches, which should match the corresponding incoming wires (see Fig. \ref{fig:FirstOrderCovariance} in Appendix \ref{App:FirstOrderOnlyProof}). Let $m'=m/\sqrt{\alpha\beta}$. In general, we obtain (to first order in $\varepsilon$, for $i=0\ldots\beta-1$, $j=0\ldots\alpha-1$):
\begin{align*}
\widehat{\psi}_+(r,l)_i &= \left(\overline{C}_+\widecheck{\psi}(r,l)\right)_i \\
&=\frac{\psi_+(r,l)}{\sqrt{\beta}}-\alpha \ii m'\varepsilon \frac{\psi_-(r,l)}{\sqrt{\alpha}} \\
&=\frac{\psi_+(r,l)-\ii m\varepsilon\psi_-(r,l)}{\sqrt{\beta}}  \\
&= \widecheck{\psi}_+(r+\varepsilon,l)_i\\
\textrm{and}\qquad\qquad\widehat{\psi}_-(r,l)_j &= \left(\overline{C}_-\widecheck{\psi}(r,l)\right)_j \\
&=\frac{\psi_-(r,l)}{\sqrt{\alpha}}-\beta \ii m'\varepsilon \frac{\psi_+(r,l)}{\sqrt{\beta}} \\
&=\frac{\psi_-(r,l)-\ii m\varepsilon\psi_+(r,l)}{\sqrt{\alpha}} \\
&= \widecheck{\psi}_-(r,l+\varepsilon)_j.
\end{align*}
Hence, the wires do match up in the first order. However, the second order cannot be fixed, even if we allow for arbitrary encodings. The proof of this statement is left for Appendix \ref{App:FirstOrderOnlyProof}.

The lack of second order covariance of the Dirac QW can be interpreted in several ways. First, as saying that the Dirac QW is not a realistic model. This interpretation motivates us to explore, in the next sections, the question whether other discrete models (QWs or QCA) could not suffer this downside, and be exactly covariant. Second, as an indication that Lorentz covariance breaks down at Planck scale. Third, as saying that we have no choice but to view $\varepsilon$ as an infinitesimal, so that we can ignore its second order. In this picture, the Dirac QW would be understood as describing an infinitesimal time evolution, but in the same formalism as that of discrete time evolutions, i.e. in an alternative language to the Hamiltonian formalism. Formulating an infinitesimal time quantum evolution in such a way has an advantage: it sticks to the language of unitary, causal operators \cite{ArrighiUCAUSAL} and readily provides a quantum simulation algorithm.

\subsection{Transformation of velocities}\label{subsec:velocities}

In Subsection \ref{subsec:DiracLorentzTransform} we defined a discrete Lorentz transform, which takes a spacetime wavefunction $\psi$ into a Lorentz transformed wavefunction $\psi'$. In Subsection \ref{subsec:focovariance} we proved that the Dirac QW is first-order covariant. Is it the case that the velocity of $\psi$ is related to the velocity of $\psi'$ according to the transformation of velocity rule of special relativity? We will show that it is indeed the case, so long as we transform the ``local velocity field'' $v(r,l)$, defined as: 
\begin{equation}
v(r,l) = \frac{|\psi_+(r,l)|^2 - |\psi_-(r,l)|^2}{||\psi(r,l)||^2} \label{Eq:LocalVelocityDiracQW}.
\end{equation}
In order to see how we arrive at this formula, let us first recall the definition of velocity in the continuous case.

For the Dirac equation, the velocity operator is obtained via the Heisenberg formula, $d \hat{x} /dt = \ii [ H, \hat{x}] = \sigma_3$ (see Eq. (\ref{eq:Dirac1D})). Thus, in the discrete setting it is natural to define the velocity operator as $\Delta X = X - W X W^\dagger$, where $X$ is the position operator, $X=\sum_x x P_x = \sum_x x \ket{x}\bra{x}$ and $W=TC$ is the walk operator. We have
\begin{align*}
\Delta X & = X - W XW^\dagger  = X - T C X C^\dagger T^\dagger =  X - T X T^\dagger \\
& = \begin{pmatrix} \sum_x xP_x & 0 \\ 0 & \sum_x xP_x \end{pmatrix} - \begin{pmatrix} \sum_x xP_{x+1} & 0 \\ 0 & \sum_x xP_{x-1} \end{pmatrix} \\
& = \sigma_3
\end{align*}
Thus the expected value of $\Delta X$ at the time slice $t=0$ is, as in the continuous case,
\begin{align}
\langle \sigma_3 \rangle_{\psi} & = \sum_{i \in \Z} |\psi_+(i,-i)|^2 - |\psi_-(i,-i)|^2 \nonumber\\
& = \sum_{i\in\Z}p(i,-i)v(i,-i) \label{Eq.GlobalVelocity}
\end{align} 
where $p(r,l) = ||\psi(r,l)||^2$. It should be noted that it is not a constant of motion. This is in fact a manifestation of the Zitterbewegung, whose connection with the continuous case was well studied by several authors \cite{StrauchPhenomena, bisio2013dirac, kurzynski2008relativistic}.
Eq. \eqref{Eq.GlobalVelocity} justifies our definition of local velocity.

Let us now consider the case of a walker which at $t=0$, $x=0$, has internal degree of freedom $\psi = (\psi_+,\psi_-)^\mathsf{T}$. We will relate $v = v(0,0)$ and $v' =  v'(0,0)$ as calculated from a Lorentz transformed observer with parameters $\alpha, \beta$. We have $v= (|\psi_+|^2 - |\psi_-|^2)/||\psi||^2$. We can deduce $|\psi_+|^2=||\psi||^2(1+v)/2$ and $|\psi_-|^2=||\psi||^2(1-v)/2$. Now, let us apply a discrete Lorentz transform. At point $(0,0)$, it takes $\psi$ into $\psi'=S\psi$, whose corresponding velocity is:
\begin{align*}
v'&=\frac{|\psi'_+|^2 - |\psi'_-|^2}{||\psi'||^2} = \frac{\alpha|\psi_+|^2 - \beta|\psi_-|^2}{\alpha|\psi_+|^2 + \beta|\psi_-|^2}\\
&= \frac{\alpha||\psi||^2(1+v) - \beta||\psi||^2(1-v)}{\alpha||\psi||^2(1+v) + \beta||\psi||^2(1-v)} \\
&= \frac{v+\frac{\alpha-\beta}{\alpha+\beta}}{1+v\frac{\alpha-\beta}{\alpha+\beta}} \\
&= \frac{v+u}{1+vu}
\end{align*}
where $u=(\alpha-\beta)/(\alpha+\beta)$ is the velocity that corresponds to the discrete Lorentz transform with parameters $\alpha, \beta$. Thus the local velocity associated to a spacetime wavefunction $\psi$ is related to the local velocity of the corresponding Lorentz transformed $\psi'$ by the rule of addition of velocities of special relativity.

\section{Formalization of Discrete Lorentz covariance in general} \label{sec:Lorentztransforms}

We will now provide a formal, general notion of discrete Lorentz transform and Lorentz covariance for Quantum Walks and Quantum Cellular Automata.

\subsection{Over Quantum Walks}\label{subsec:LorentztransformsQWs}

 Beforehand, we need to explain which general form we assume for Quantum Walks.
\subsubsection{General form of Quantum Walks} 
\begin{figure}
\includegraphics[width=\columnwidth]{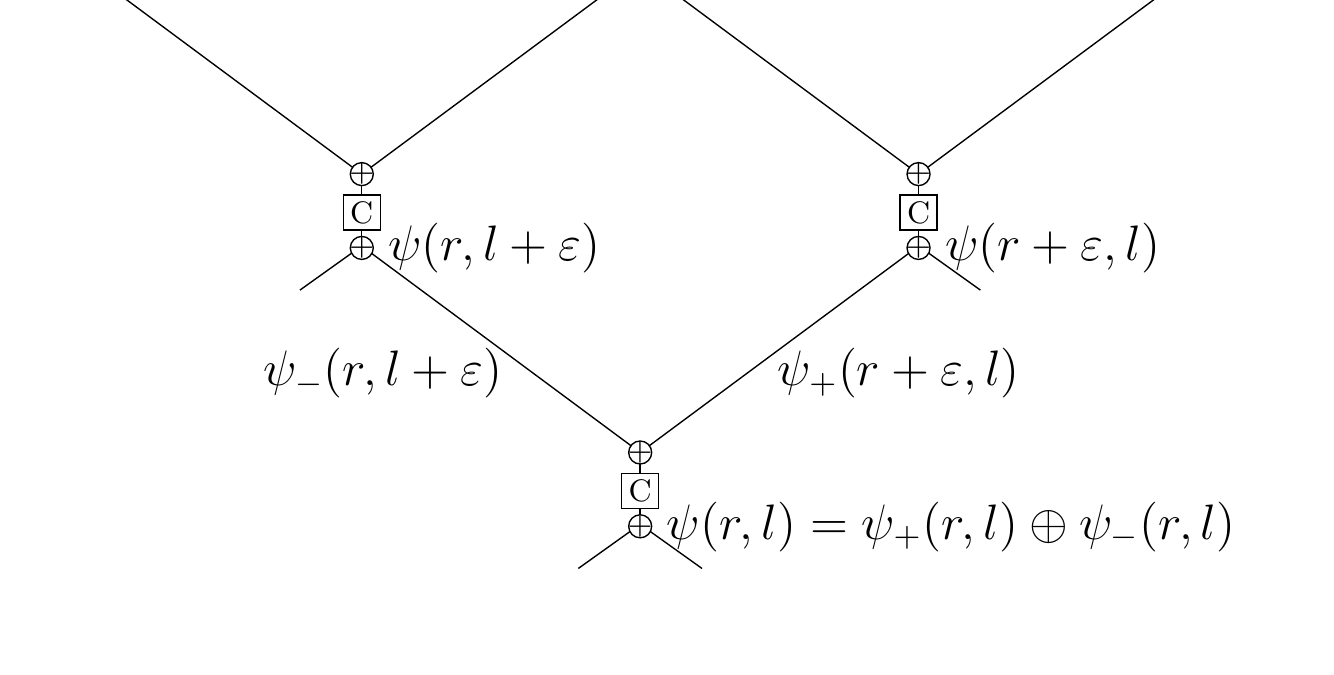}
\caption{\label{fig:QWGeneral} {\em The circuit for a general QW.} The wire dimension is $d$, meaning that $\psi_+(r,l)=(\psi^1_+(r,l),\ldots,\psi^d_+(r,l))^\mathsf{T}$, etc.}
\end{figure}

Intuitively speaking, a Quantum Walk (QW) is a single particle or walker moving in discrete-time steps on a lattice. Axiomatically speaking, QWs are shift-invariant, causal, unitary evolutions over the space $\bigoplus_{\mathbb{Z}} {\cal H}_{c}$, where $c$ is the dimension of the internal degrees of freedom of the walker. Constructively speaking, in turns out \cite{GrossNesmeVogtsWerner} that, at the cost of some simple recodings, any QW can be put in a form which is similar to that of the circuit for the Dirac QW shown Fig. \ref{fig:conventions}$(c)$. In general, however, $c$ may be larger than $2$ (the case $c$ equal $1$ is trivial \cite{MeyerQLGI}). But it can always be taken to be even, so that the general shape for the circuit of a QW can be expressed as in Fig. \ref{fig:QWGeneral}. Notice how, in this diagram, each wire carries a $d$-dimensional vector $\psi_\pm(r,l)$. We will say that the QW has `wire dimension' $d$. Incoming wires get composed together with a direct sum, to form  a $2d$-dimensional vector $\psi(r,l)$. The state $\psi(r,l)$ undergoes a $2d\times2d$ unitary gate $C$ to become some $\psi'(r,l)=\psi_+'(r+\varepsilon,l)\oplus \psi_-'(r,l+\varepsilon)$, etc. The unitary gate $C$ is called the `coin'. Algebraically speaking, this means that a QW can always be assumed to be of the form:
\begin{equation*}
\diag{e^{\varepsilon\partial_r}\Id_d}{e^{\varepsilon\partial_l}\Id_d}{\psi}=C{\psi}.
\end{equation*}

\subsubsection{Lorentz transforms for QW} 
The formalization of a general notion of Lorentz transform for QWs generalizes that presented in Section \ref{sec:aLorentz transform}. Consider a QW having wire dimension $d$, and whose $2d\times 2d$ unitary coin is $C_m$, where the $m$ are parameters. A Lorentz transform $L_{\alpha,\beta}$ is specified by:
\begin{itemize}
\item a function $m'=f_{\alpha,\beta}(m)$, such that $f_{\alpha'\alpha, \beta'\beta}=f_{\alpha',\beta'}\circ f_{\alpha,\beta} $.
\item a family of isometries $E_\alpha$ from ${\cal H}_{d}$ to $\bigoplus_\alpha{\cal H}_{d}$, such that $(\bigoplus_\alpha E_{\alpha'})E_\alpha = E_{\alpha'\alpha}$.
\end{itemize}
Above we used the notation $\bigoplus_\alpha{\cal H}_d = \bigoplus_{i=1\ldots \alpha} {\cal H}_d$.
Consider ${\psi}$ a spacetime wavefunction (at this stage it is not necessary to assume that it is a solution of the QW). Switching to lightlike coordinates, its Lorentz transform $\psi'=L_{\alpha,\beta}\psi$ is obtained by:
\begin{itemize}
\item for every $(r,l)$, computing:  $\widecheck{\psi}_+=E_\beta\psi_+$, $\widecheck{\psi}_-=E_\alpha\psi_-$, and $\widecheck{\psi}=\widecheck{\psi}_+\oplus\widecheck{\psi}_-=\overline{E} {\psi}$.
\item for every $(r,l)$, replacing: the point $(r,l)$ by the lightlike $\alpha\times\beta$ rectangular patch of spacetime 
\begin{align}
\left(\overline{C}_{m'}(i,j)\widecheck{\psi}(r,l)\right)_{i=0\ldots\alpha-1,j=0\ldots\beta-1}\label{eq:genLorentz transform}
\end{align}
with $\overline{C}_{{m}'}(i,j)$ as in Remark \ref{rk:detregion} and Fig. \ref{fig:determined}.
\end{itemize}
Again, Fig. \ref{fig:Lorentz transform} illustrated an example of such a transformation.

\subsubsection{Lorentz covariance for QW}
The formalization of a general notion of Lorentz covariance for QWs generalizes that presented in Subsection \ref{subsec:focovariance}. Consider a QW having wire dimension $d$ whose $2d\times 2d$ and unitary coin unitary coin $C_m$, where the $m$ are parameters. Consider $\psi$ a spacetime wavefunction which is a solution of this QW. We just gave the formalization of a discrete general notion of Lorentz transform taking a spacetime wavefunction $\psi$ into another spacetime wavefunction $\psi'=L_{\alpha,\beta}\psi$, and parameters ${m}$ into ${m}'$. Is it the case, for any $\alpha$ and $\beta$, that the spacetime wavefunction $\psi'$ is a solution of the same QW, but with parameters ${m}'$? If so, the QW is said to be covariant with respect to the given discrete Lorentz transform. Now, the above-defined discrete Lorentz transform is obtained by replacing each point with a lightlike $\alpha\times\beta$ rectangular patch of spacetime, which, by definition, is internally a piece of spacetime solution of the Dirac QW see Eq. \eqref{eq:genLorentz transform}. But again, is it the case that the patches match up to form the entire spacetime wavefunction of a solution? Let us again define
\begin{equation*}
\widehat{\psi}_+(r,l)=(\overline{C}_{m'})_+\widecheck{\psi}(r,l)\quad\textrm{and}\quad \widehat{\psi}_-(r,l)=(\overline{C}_{m'})_-\widecheck{\psi}(r,l).
\end{equation*}
We need: $\widehat{\psi}_+(r,l)=\widecheck{\psi}_+(r+\varepsilon,l)$ and $\widehat{\psi}_-(r,l)=\widecheck{\psi}_-(r,l+\varepsilon)$.
An equivalent, algebraic way of stating these two requirements is obtained as follows:
\begin{align*}
\widecheck{\psi}_+(r+\varepsilon,l) \oplus \widecheck{\psi}_-(r,l+\varepsilon) = \widehat{\psi}_+(r,l) \oplus \widehat{\psi}_-(r,l) \nonumber
\end{align*}
Equivalently,
\begin{align}
& \left(E_\beta\oplus E_\alpha  \right)\left(\psi_+(r+\varepsilon,l)\oplus \psi_-(r,l+\varepsilon)\right) = \nonumber \\ 
& \left(\overline{C}_{{m}'}(\alpha,\cdot)\oplus \overline{C}_{{m}'}(\cdot,\beta)\right)\left(E_\beta\oplus E_\alpha  \right){\psi}(r,l)\nonumber\\ 
\Leftrightarrow  & \left(E_\beta\oplus E_\alpha  \right)C_m \psi(r,l) = \overline{C}_{{m}'}\left(E_\beta\oplus E_\alpha \right){\psi}(r,l) \nonumber\\ 
\Leftrightarrow & \left(E_\beta\oplus E_\alpha  \right)C_m = \overline{C}_{{m}'}\left(E_\beta\oplus E_\alpha  \right)\nonumber\\
\Leftrightarrow & \overline{E} C_m = \overline{C}_{{m}'}\overline{E}. \label{eq:consistency}
\end{align}

This expresses discrete Lorentz covariance elegantly, as a form of commutation relation between the evolution and the encoding. Diagrammatically this is represented by Fig. \ref{fig:CovRules}$(a)$. The isometry of the $E_\alpha$ can also be represented diagrammatically, cf. Fig. \ref{fig:CovRules}$(b)$. Combining both properties straightforwardly leads to 
\begin{align}
 C_m &= \overline{E}^\dagger \overline{C}_{{m}'} \overline{E}.\label{eq:pointconsistency}
\end{align}
This is represented as Fig. \ref{fig:OtherRules}$(a)$, which of course can be derived diagrammatically from Fig. \ref{fig:CovRules}. Is this diagrammatic theory powerful enough to be considered an abstract, pictorial theory of Lorentz covariance, in the spirit of \cite{CoeckeBoxes}?

\subsubsection{Diagrammatic Lorentz covariance for QW} 
\begin{figure*}
\includegraphics[scale=.7]{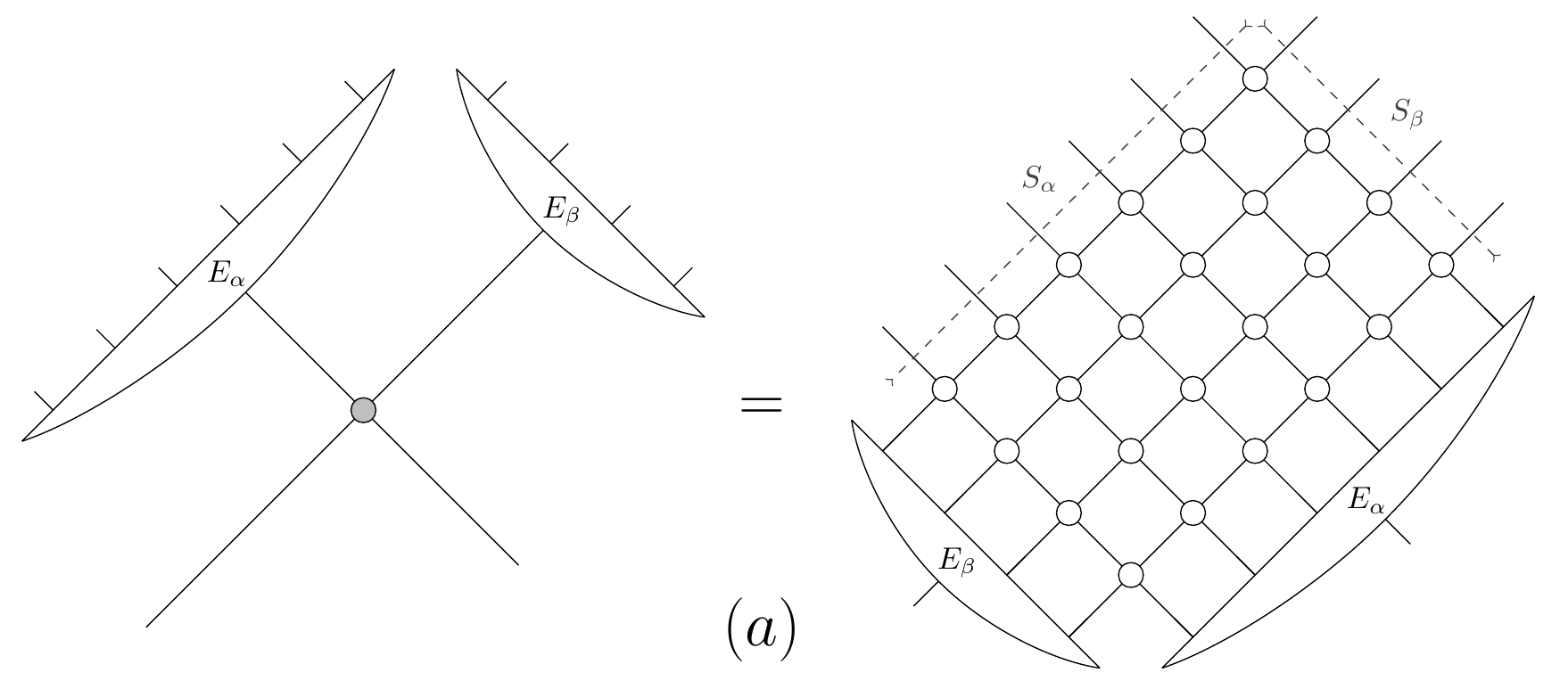}
\includegraphics[scale=.7]{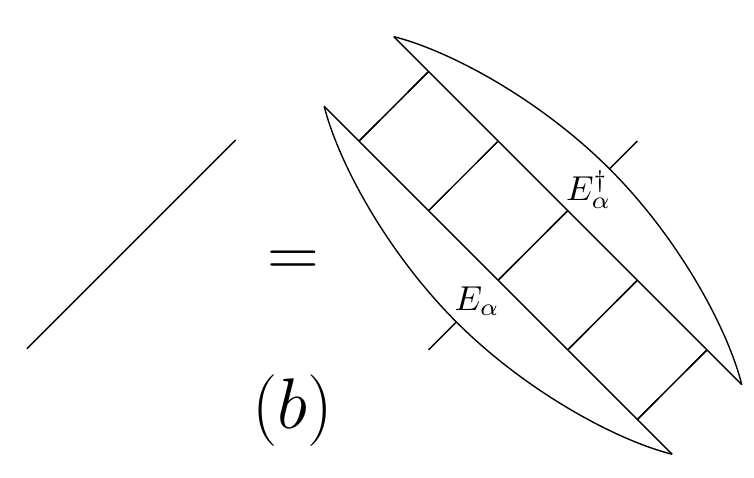}
\caption{\label{fig:CovRules} {\em Basic covariance rules}. $(a)$ Expresses the fundamental covariance condition of Eqs. \eqref{eq:consistency} and \eqref{eq:consistency2}. The dashed line is optional, it is an indication which results from using this rule: it tells us that the state of these wires belongs to the subspace $S_\alpha$. The gray and white dots stand for the same unitary interaction, but with different parameters. $(b)$ Expresses the isometry of the encodings used for the discrete Lorentz transform.}
\end{figure*}

Combining the diagrammatic equalities of Fig. \ref{fig:CovRules}, we can almost rewrite the spacetime circuit of a QW with coin $C_m$, into its Lorentz transformed version, for any parameters $\alpha,\beta$\ldots but not quite. A closer inspection shows that this can only be done over regions such as past cones, by successively: 1/ Introducing pairs of encodings via rule Fig. \ref{fig:CovRules}$(b)$ along the border of the past cone; 2/ Pushing back towards the past the bottom $E$ via rule Fig. \ref{fig:CovRules}$(a)$, thereby unveiling the Lorentz transformed past cone. Whilst this limitation to past-cone-like regions may seem surprising at first, there is a good intuitive reason for that. Indeed, the diagrammatic equalities of Fig. \ref{fig:CovRules} tell you that you can locally zoom into a spacetime circuit; but you can only locally zoom out if you had zoomed in earlier, otherwise there may be a loss of information. This asymmetry is captured by the fact that Fig. \ref{fig:CovRules}$(a)$ cannot be put upside-down, time-reversed. It follows that you should not be able to equalize an entire spacetime circuit with its complete Lorentz transform, at least not without using further hypotheses. And indeed, when we local Lorentz transform an entire past cone, its border is there to keep track of the fact that this region was locally zoomed into, and that we may later unzoom from it, if we want.\\
\begin{figure}
\includegraphics[scale=.6]{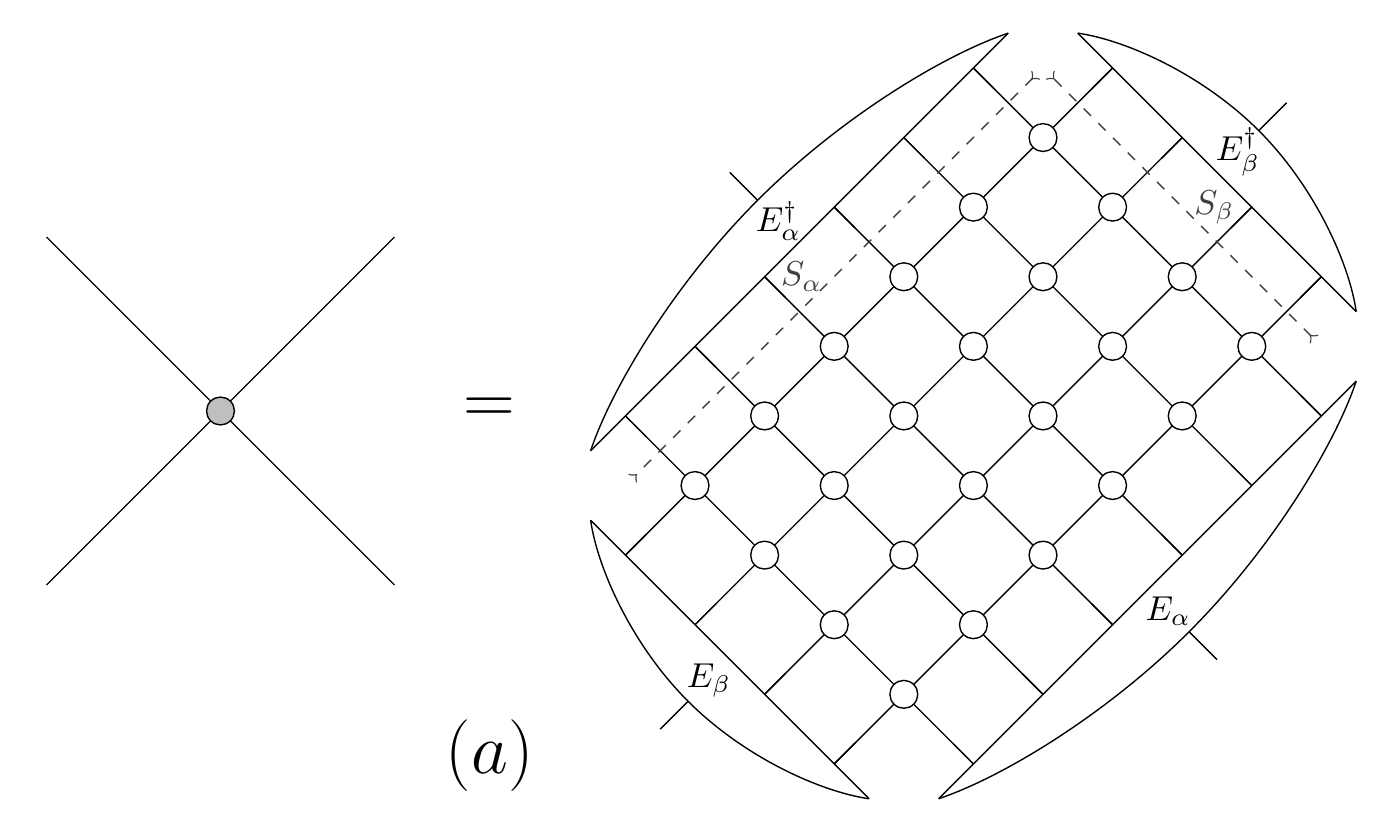}
\includegraphics[scale=.6]{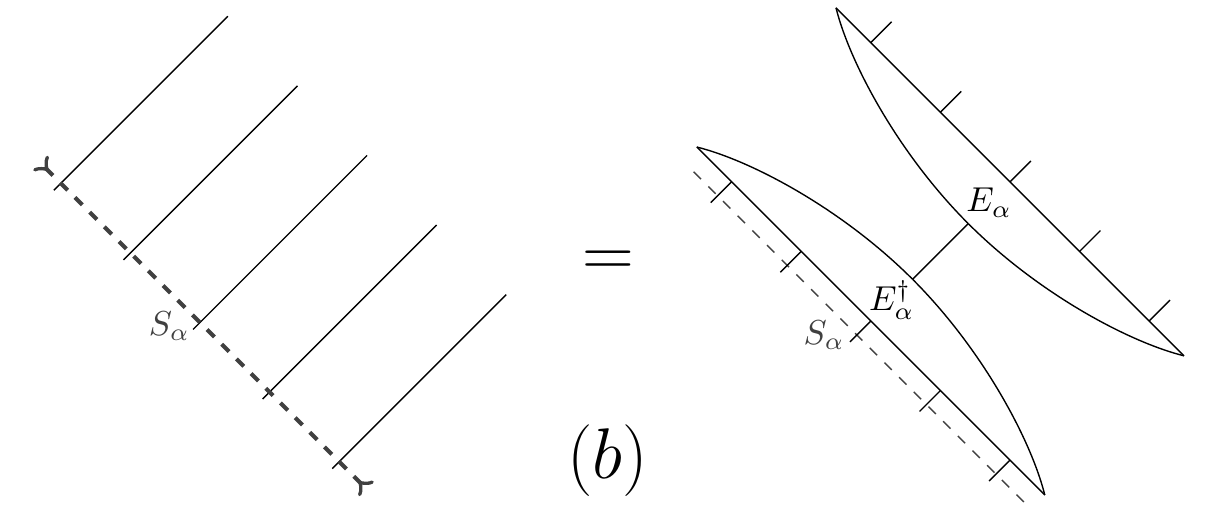}
\caption{\label{fig:OtherRules} {\em Completed covariance rules}. $(a)$ is a theorem, derived from the diagrams of Fig. \ref{fig:CovRules}, see also Eqs \eqref{eq:pointconsistency} and \eqref{eq:pointconsistency2}. It expresses the idea of a Lorentz transform being a zoom in. The dashed line is optional, it is an indication which results from using this rule: it tells us that the state of these wires belongs to the subspace $S_\alpha$. The gray and white dots stand for same unitary interaction, but with different parameters. $(b)$ is a conditional rule: the thicker dashed line is a precondition for the equality to hold. Again it follows from the isometry of the encodings used for the discrete Lorentz transform, see also Eq. \eqref{eq:subspace}.}
\end{figure}
\begin{figure*}
\includegraphics[width=\textwidth]{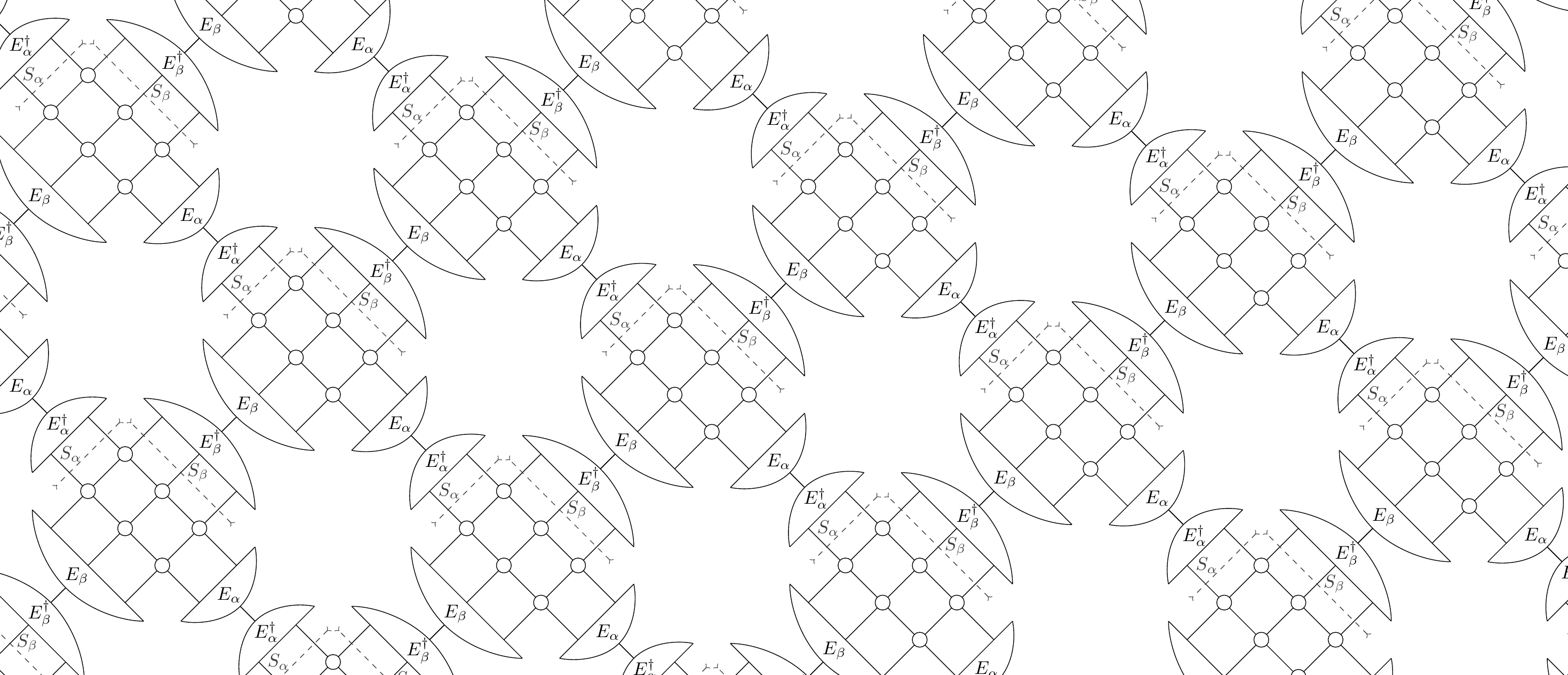}
\caption{\label{fig:transformingall} {\em Performing a complete Lorentz transform via the completed covariance rules}. A Lorentz transform with parameters $\alpha=2$ and $\beta=3$ consists in replacing each point by a $2\times 3$ rectangular patch of spacetime using isometric encodings.}
\end{figure*}
Now, could we add a further diagrammatic rule which would allow us to perform an complete Lorentz transformation, perhaps at the cost of annotating our spacetime circuit diagrams with information on whom has been zoomed into? Those annotations are the dashed lines of Fig. \ref{fig:CovRules} and Fig. \ref{fig:OtherRules}$(a)$. Clearly, as we use those rules, we know whether some bunch of wires lives in the subspace $S_{\alpha}$ of the projector $E_\alpha E_\alpha^\dagger$, and we can leave that information behind. Moreover, on this subspace, it is the case that 
\begin{align}
E_\alpha E_\alpha^\dagger=\Id_{S_{\alpha}}.\label{eq:subspace}
\end{align}
Then, representing this last equation in rule Fig. \ref{fig:OtherRules}$(b)$, which is conditional on the annotation being there (the other rules are non-conditional, they provide the annotations), we reach our purpose. Indeed, in order to perform a 
complete Lorentz-transform we can now apply the rule Fig. \ref{fig:OtherRules}$(a)$ everywhere, leading to Fig. \ref{fig:transformingall}, and then remove the encoding gates everywhere via Fig. \ref{fig:OtherRules}$(b)$. 
Thus, it could be said that the rewrite rules of Fig. \ref{fig:OtherRules} provide an abstract, pictorial theory of Lorentz covariance. They allow to equalize, spacetime seen by a certain observer, with spacetime seen by another, inertial observer. Besides their simplicity, the local nature of the rewrite rules is evocative of the local Lorentz covariance of General Relativity. This is explored a little further in Subsection \ref{subsec:nonhomog}.

\subsubsection{Inverse transformations and equivalence upon rescaling} 
In analogy with the continuum case, we would like the inverse of a Lorentz transform $L_{\alpha,\beta}$ to be $L_{\beta,\alpha}$, i.e.
\begin{equation}\label{eq:inverselorentz}
L_{\alpha,\beta}L_{\beta,\alpha} = \Id.
\end{equation}
However, according our definitions of $L_{.,.}$, we know that $L_{\alpha,\beta}L_{\beta,\alpha}$ is a transformation such that
\begin{itemize}
\item each point $(r,l)$ is replaced by the lightlike $\alpha\beta\times\alpha\beta$ square patch of spacetime, with left-incoming wires $F\psi_+(r,l)$, right-incoming wires $F\psi_-(r,l)$,  right-outgoing wires $F\psi_+(r+\varepsilon,l)$ and left-outgoing wires $F\psi_-(r,l+\varepsilon)$, where 
\begin{equation*}
F = \left(\bigoplus_\alpha E_\beta\right)E_\alpha = E_{\beta\alpha} = E_{\alpha\beta} = \left(\bigoplus_\beta E_\alpha\right)E_\beta
\end{equation*}
\item the coin parameter $m$ is mapped to $m'=f_{\alpha\beta,\alpha\beta}(m)$.
\end{itemize}
Hence, if we are to claim \eqref{eq:inverselorentz} we need to identify any two spacetime diagrams which satisfy these relations. This is achieved as a special case in the completed diagrammatic theory of Fig. \ref{fig:OtherRules}.

\subsection{Over Quantum Cellular Automata} \label{subsec:LorentztransformsQCA}

\subsubsection{General form of Quantum Cellular Automata} 
\begin{figure}
\includegraphics[width=\columnwidth]{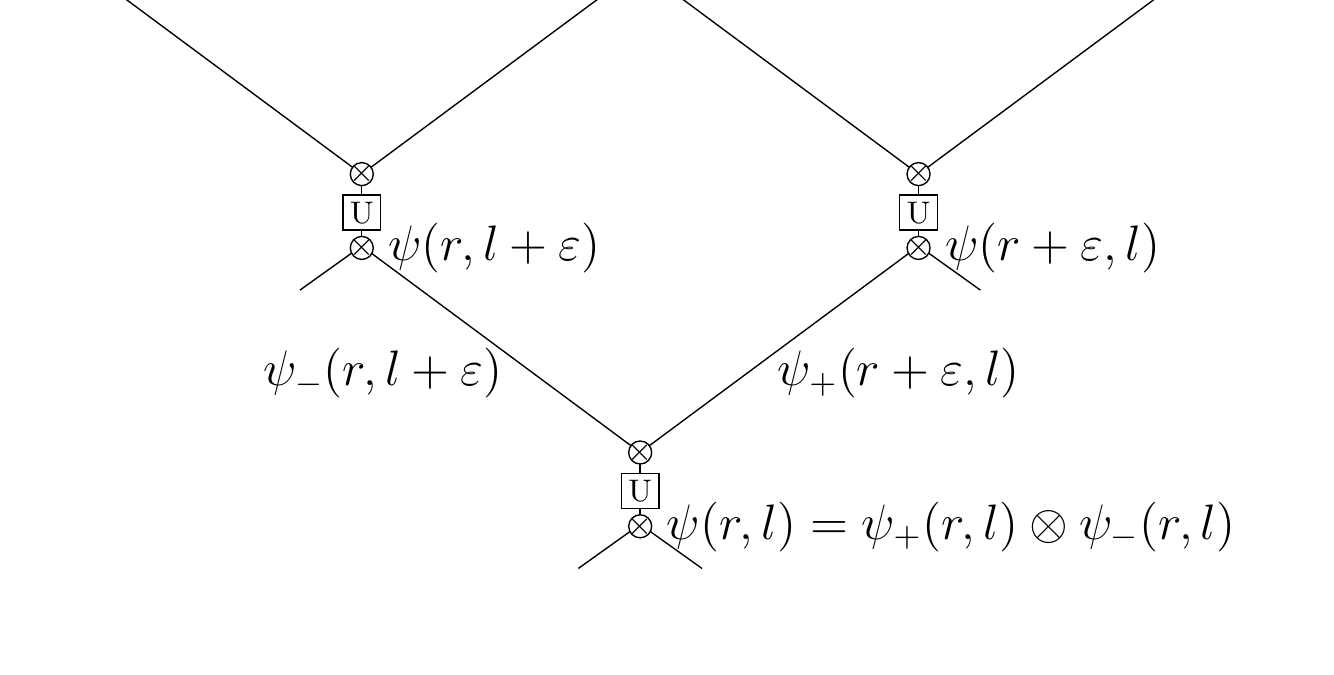}
\caption{\label{fig:QCAGeneral} {\em The circuit for a general QCA.}}
\end{figure}

Intuitively speaking, a Quantum Cellular Automata (QCA) is a multiple walkers QW. The walkers may or may not interact, their numbers may or may not be conserved. Axiomatically speaking, a QCA is a shift-invariant, causal, unitary evolution over the space $``\bigotimes_{\mathbb{Z}} {\cal H}_{c}"$, where $c$ is the dimension of the internal degrees of freedom of each site. Actually, care must be taken when defining such infinite tensor products, but two solutions exist \cite{SchumacherWerner,ArrighiLATA,ArrighiIJUC}.  Constructively speaking, it turns out \cite{SchumacherWerner,ArrighiLATA,ArrighiIJUC} that, at the cost of some simple recodings, any QCA can be put in the form of a quantum circuit. This circuit can then be simplified \cite{ArrighiPQCA} to bear strong resemblance with the circuit of a general QW seen in Fig. \ref{fig:QWGeneral}. In particular $c$ can always be taken to be $d^2$, so that the general shape for the quantum circuit of a QCA is that of Fig. \ref{fig:QCAGeneral}. Notice how, in this diagram, each wire carries a $d$-dimensional vector $\psi_\pm(r,l)$. We will say that the QCA has `wire dimension' $d$. Incoming wires get composed together with a tensor product, to form  a $d^2$-dimensional vector $\psi(r,l)$. The state $\psi(r,l)$ undergoes a $d^2\times d^2$ unitary gate $U$ to become some $\psi_+(r+\varepsilon,l)\otimes \psi_-(r,l+\varepsilon)$, etc. The unitary gate $U$ is called the `scattering operator'. Notice how, to some extent, the QCA are alike QW up to replacing $\oplus$ by $\otimes$.  Algebraically speaking, the above means that one time-step of a QCA can always be assumed to be of the form:
\begin{align*}
{\psi}\mapsto 
&\left(\bigotimes_{2\mathbb{Z}+1}U\right)\left(\bigotimes_{2\mathbb{Z}}U\right){\psi}.
\end{align*}

\subsubsection{Lorentz transforms for QCA}
The formalization of a general notion of Lorentz transform for QCA is obtained from that over QW essentially by changing occurrences of $\oplus$ into $\otimes$. Indeed, consider a QCA having wire dimension $d$, and whose $d^2\times d^2$ unitary scattering operator $U$ has parameters $m$. A Lorentz transform $L_{\alpha,\beta}$ is specified by:
\begin{itemize}
\item a function ${m}'=f_{\alpha,\beta}({m})$ such that $f_{\alpha'\alpha, \beta'\beta}=f_{\alpha',\beta'}\circ f_{\alpha,\beta} $.
\item a family of isometries $E_{\alpha}$ from ${\cal H}_{d}$ to $\bigotimes_\alpha {\cal H}_{d}$, such that $(\bigotimes_\alpha E_{\alpha'})E_\alpha = E_{\alpha'\alpha}$.
\end{itemize}
There is a crucial difference with QWs, however, which is that we cannot easily apply this discrete Lorentz transform to a spacetime wavefunction. Indeed, consider ${\psi}$ a spacetime wavefunction. For every time $t$, the state $\psi(t)$ may be a large entangled state across space. What meaning does it have, then, to select another spacelike surface? What meaning does it have to switch to lightlike coordinates? Unfortunately the techniques which were our point of departure for QWs, no longer apply. Fortunately, the algebraic and diagrammatic techniques which were out point of arrival for QWs, apply equally well to QCA, so that we may still speak of Lorentz-covariance. 

\subsubsection{Lorentz covariance for QCA}
Again, the formalization of the notion of Lorentz-covariance for QCA cannot be given in terms of $\psi'$ being a solution if $\psi$ was a solution, because we struggle to speak of $\psi'$. Instead, we define Lorentz-covariance straight from the algebraic view:
\begin{align}
\left(E_\beta\otimes E_\alpha  \right)U_m &= \overline{U}_{{m}'}\left(E_\beta\otimes E_\alpha  \right)\nonumber\\
\textrm{i.e. }\quad\overline{E} U_m &= \overline{U}_{{m}'}\overline{E}.\label{eq:consistency2}
\end{align}
Diagrammatically this is represented by the same figure as for QWs, namely Fig. \ref{fig:CovRules}$(a)$. The isometry of the $E_\alpha$ is again represented by Fig. \ref{fig:CovRules}$(b)$. Algebraically speaking, combining both properties again leads to 
\begin{align}
 U_m &= \overline{E}^\dagger \overline{U}_{{m}'} \overline{E}.\label{eq:pointconsistency2}
\end{align}
Which diagrammatically this is again represented as Fig. \ref{fig:OtherRules}$(a)$. For the same reasons, the conditional rule Fig. \ref{fig:OtherRules}$(b)$ again applies: the whole diagrammatic theory carries through unchanged from QWs to QCA.

\subsection{Non-homogeneous discrete Lorentz transforms and non-inertial observers}\label{subsec:nonhomog}

\begin{figure*}
\includegraphics[width=\textwidth]{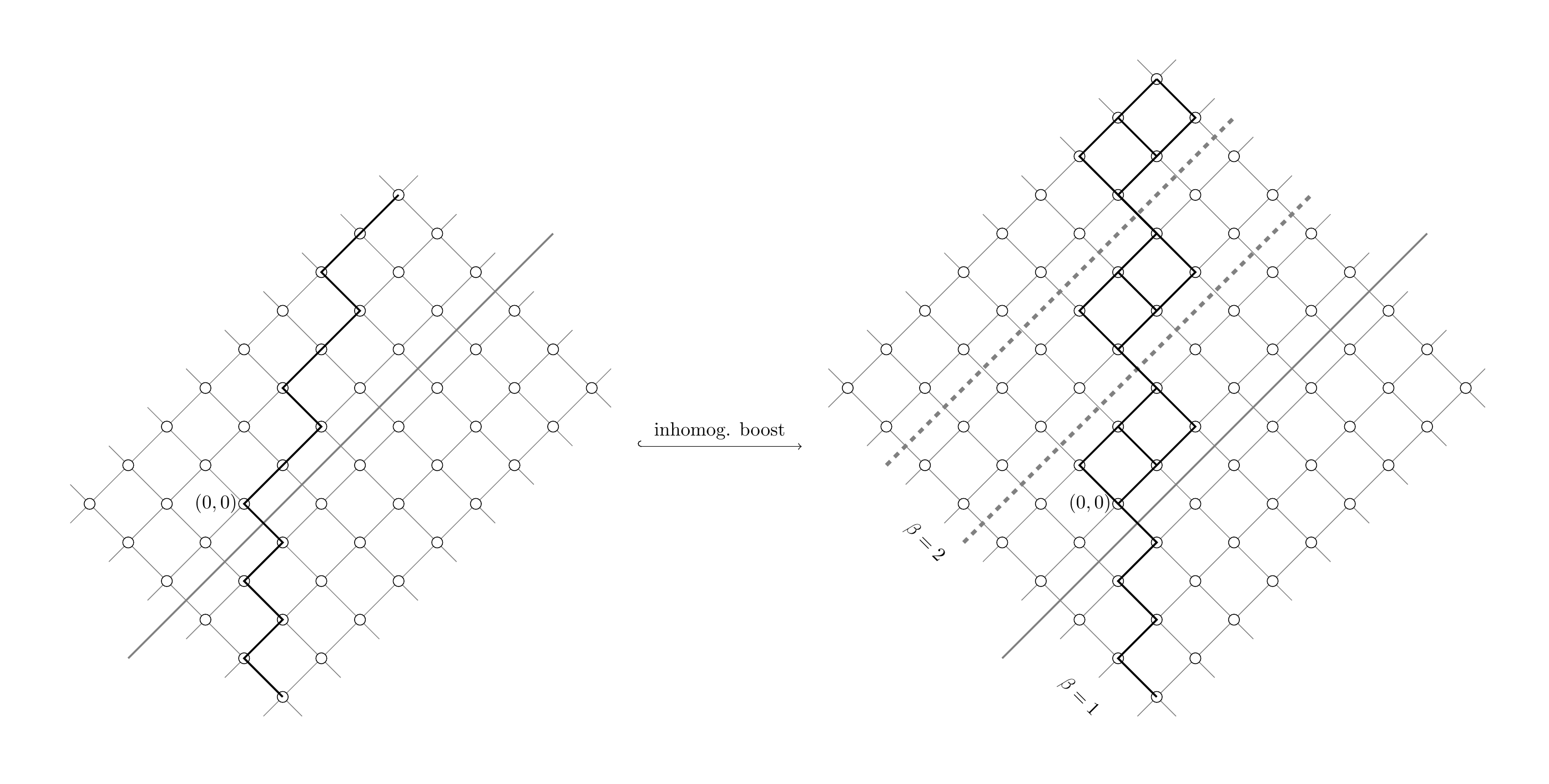}
\caption{\label{fig:nonhomogeneoussimple} {\em An inhomogeneous transformation for a non-inertial observer.} The region above the gray line undergoes a Lorentz transform with parameters $\alpha=1$ and $\beta=2$, whilst the region below is left unchanged. After the inhomogeneous transformation, the observer is at rest.}
\end{figure*}

Nothing in the above developed diagrammatic theory forbids us to apply different local discrete Lorentz transforms to different points of spacetime, so long as point $(r,l)$ and point $(r+\varepsilon,l)$ (resp. point $(r,l+\varepsilon)$) have the same parameter $\beta$ (resp. $\alpha$). This constraint propagates along lightlike lines, so that there can be, at most, one different $\alpha_r$ (resp. $\beta_l$) per right-moving (resp. left-moving) lightlike line $r$ (resp. $l$). We call this a non-homogeneous discrete Lorentz transform of parameters $(\alpha_r)$, $(\beta_l)$.

The circuit which results from applying such a non-homogeneous discrete Lorentz transform is, in general, a non-homogeneous QWs (resp. QCA), as it may lack shift-invariance in time and space. This is because the coin $C_m$ (resp. scattering unitary $U_m$) of the point $(r,l)$ gets mapped into lightlike $\alpha_r\beta_l$-rectangular patch of spacetime $\overline{C}_{m'}$ (resp. $\overline{U}_{m'}$), whose parameters $m'=f_{\alpha_r,\beta_l}(m)$ may depend, in general, upon the position $(r,l)$. This problem is avoided if $f_{\alpha\beta}=f$ does not depend upon $\alpha$ and $\beta$, as in the example which will be introduced in Section \ref{sec:ClockQCA}.

Provided that the condition $f_{\alpha\beta}=f$ is met, we can now transform between non-inertial observers by a non-homogeneous discrete Lorentz transform. Figure \ref{fig:nonhomogeneoussimple} illustrates this with the simple example of an observer which moves one step right, one step left, until it reaches point $(0,0)$ where it gets accelerated, and continues moving two steps right, one step left etc. We choose $\beta_{l}=1$ for $l < 0$,$\beta_{l}=2$ for $l\geq 0$ and $\alpha_{r}=1$ for all $r$. This has the effect of slowing down the observer just beyond the point $(0,0)$. All along his trajectory, he now has to move two steps right for every two steps left that he takes, so that he is now at rest.

In general, suppose that an observer moves $a_k$ steps to the right, $b_k$ steps left, $a_{k+1}$ steps right, etc. He does this starting from position $r_k = r_{k-1}+a_k$ and $l_k=l_{k-1}+b_k$.
For every $k$, let $M_k$ be the least common multiple of $a_k$ and $b_k$. We choose $\alpha_r=M_k/a_k$ for $r_{k-1} \leq r < r_{k}$ and $\beta_l=M_k/b_k$ for $l_{k-1} \leq l < l_{k}$. Let us perform the non-homogeneous discrete Lorentz transform of parameters $(\alpha_r)$, $(\beta_l)$. Then, the observer now moves $M_k$ steps right for every $M_k$ steps left he takes, and then $M_{k+1}$ steps right for every $M_{k+1}$ steps left, etc.

\section{The Clock QW}\label{sec:ClockQW}

Equipped with a formal, general notion of Lorentz transform and Lorentz covariance for QW, we can now seek for an exactly covariant QW.

\subsection{Definition}

\def\rs{p}
\def\ls{q}

In the classical setting, covariance of random walks has already been explored \cite{wall1988discrete}. The random walk of \cite{wall1988discrete} uses a fair coin, but is nonetheless biased in the following way: after a (fair) coin toss the walker moves during $\rs$ time steps to the right (resp. during $\ls$ time steps to the left). There is a reference frame in which the probability distribution is symmetric, namely that with velocity $u = (\rs - \ls)/(\rs + \ls)$. Changing the parameters $\rs$ and $\ls$ corresponds to performing a Lorentz transform of the spacetime diagram. 

Now we will make an analogous construction in the quantum setting. The main point is to enlarge the coin space so that the coin operator is idle during $\rs$, or $\ls$, time steps. The coin space will be ${\mathcal{H}}_C = \mathcal{H}_C^+ \oplus \mathcal{H}_C^-$, where $\mathcal{H}_C^+ \cong \mathcal{H}_C^- = \ell^2(\mathbb Q^{\geq 0})$. The Hilbert space of the quantum walk is then ${\mathcal{H}} = \ell^2(\Z) \otimes {\mathcal{H}_C}$, whose basis states will be indicated by $\ket{x,h^s}$, with $h\in\mathbb Q^{\geq0}$, $s=\pm$.

This $\mathcal{H}_C^\pm$ will act as a ``counter''. When $h>0$, the walker moves without interaction and the counter is decreased. When the counter reaches $0$, the effective coin operator is applied and the counter is reset.

The evolution of the Clock QW with parameters $\rs,\ls$ is defined on the subspace $\mathcal{H}_C^{\rs,\ls}$ of $\mathcal{H}_C$ spanned by the $\rs  + \ls$ vectors $\{\ket{\frac{i}{\rs}^+},\ket{\frac{j}{\ls}^-}\}$ with $i=0,\dots,\rs - 1$ and $j=0,\dots,\ls - 1$, as follows:
\begin{widetext}
\begin{equation}
W_{\rs,\ls}\ket{x,h^s} = \left\{
\begin{array}{ll}
  \ket{x+1,(h-\frac1{\rs})^+} & \mbox{for } s=+, \quad 0 < h \leq 1-\frac1{\rs}\\
  \ket{x-1,(h-\frac1{\ls})^-} & \mbox{for } s=-, \quad 0 < h \leq 1-\frac1{\ls}\\
  a \ket{x+1,(1-\frac1{\rs})^+} + b \ket{x-1,(1-\frac1{\ls})^-} & \mbox{for } s=+, \quad h=0\\
  c \ket{x+1,(1-\frac1{\rs})^+} + d \ket{x-1,(1-\frac1{\ls})^-}  & \mbox{for } s=-, \quad h=0 \\
\end{array} \right. \label{eq:ClockQW_Map}
\end{equation}
\end{widetext}

This map is unitary provided that the $2\times 2$ matrix $C$ of coefficients $C_{11}=a$, $C_{12}=b$, $C_{21}=c$, $C_{22}=d$ is unitary. For instance we could choose, as for the Dirac QW, $a=d=\cos(m \varepsilon)$, $b=c=-\ii \sin(m \varepsilon)$. 

The Clock QW with parameters $\rs$ and $\ls$ will only be used over $\ell^2(\Z)\otimes \mathcal{H}_C^{\rs,\ls}$ where it admits a matrix form which we now provide (over the rest of $\mathcal{H}_C$ it can be assumed to be the identity). From Eq. (\ref{eq:ClockQW_Map}) we can write $W_{\rs,\ls} = T_{\rs,\ls}C_{\rs,\ls}$ where $T_{\rs,\ls}$ is the shift operator,
\begin{equation*}
T_{\rs,\ls} = \text{diag} \left(\overbrace{e^{-\varepsilon \partial_x},\dots,e^{-\varepsilon \partial_x}}^{\rs\mbox{ \footnotesize times}},\overbrace{e^{\varepsilon \partial_x},\dots,e^{\varepsilon \partial_x}}^{\ls\mbox{ \footnotesize times}} \right)
\end{equation*}
and $C_{\rs,\ls}$ is the coin operator:
\begin{equation*}
C_{\rs,\ls} = \left(\begin{array}{cc|cc}
0 & \Id_{\rs-1} & 0 & 0 \\
a & 0 & b & 0 \\
\hline
0 & 0 & 0 & \Id_{\ls - 1} \\
c & 0 & d & 0
\end{array}\right)
\end{equation*}

Hence, the Clock QW has an effective coin space of finite dimension $\rs+\ls$. However, we will see that this dimension changes under Lorentz transforms.

\subsection{Covariance}

\begin{figure*}
  \subfloat {
    \raisebox{3cm}{\includegraphics[scale=0.5]{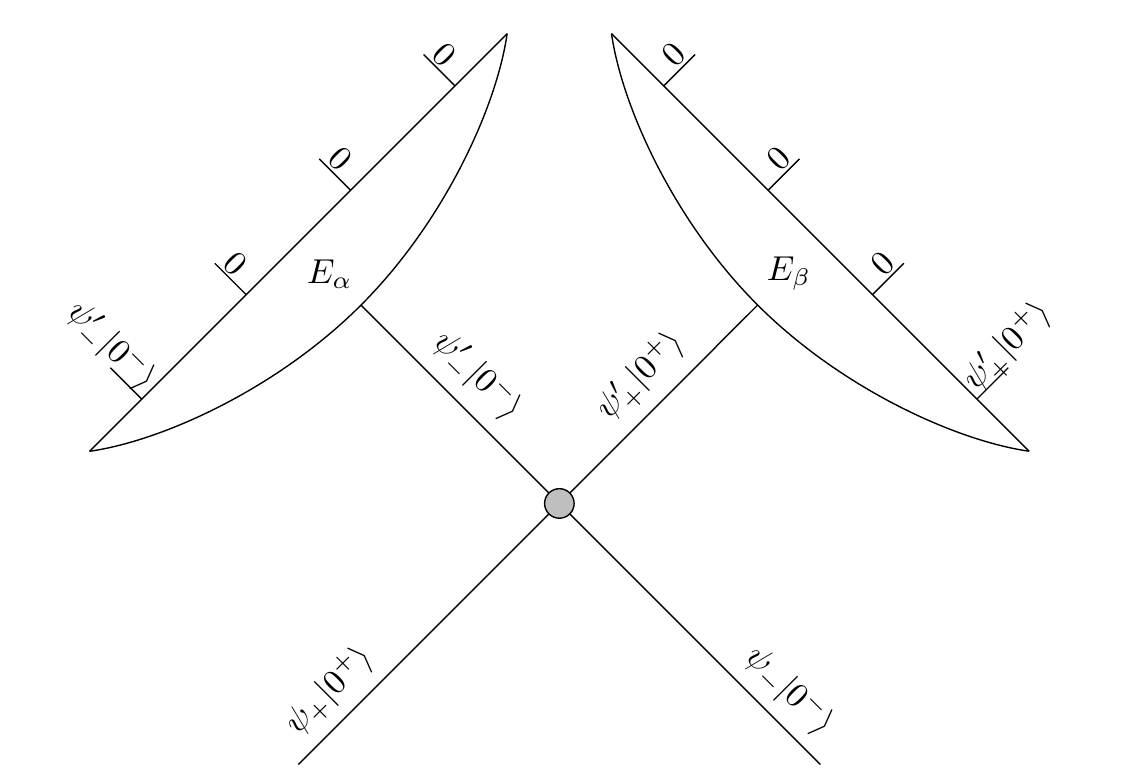}}
  }
  \raisebox{4.5cm}{\large =}
  \subfloat {
    \includegraphics[scale=0.5]{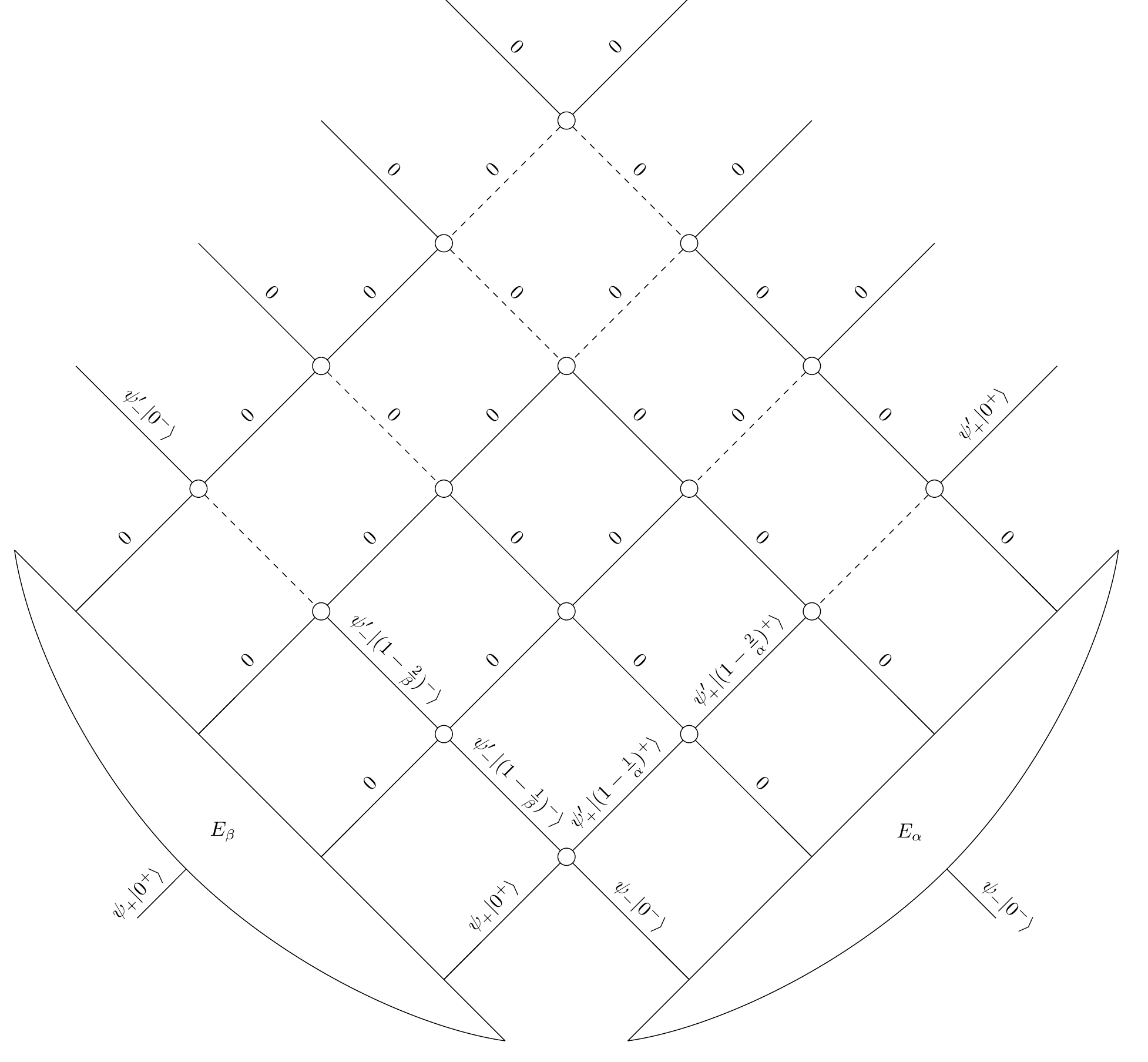}
  }
  \caption{\label{fig:ClockQWCov} {\em Covariance of the Clock QW.} This is the transformation given by $\alpha,\beta$ of the Clock QW with parameters $\rs=1,\ls=1$.}
\end{figure*}

In order to prove covariance, we need to find isometries satisfying the equation expressed by Fig. \ref{fig:CovRules}$(a)$. Let us consider isometries $E_{\alpha}: \mathcal{H}_C \to \bigoplus_{\alpha} \mathcal{H}_C$ defined by:
\begin{align*}
E_{\alpha} \ket{h^s} = (\ket{h^s} \oplus \overbrace{0 \oplus \dots \oplus 0}^{\mbox{$\alpha-1$ times}}) \\
\end{align*}
(the Hilbert spaces in the direct sum are ordered from the bottom wire to the top one, as in remark \ref{rk:detregion}). In Fig. \ref{fig:ClockQWCov} it is proved that this choice actually satisfies the covariance relation $\overline{E}C_{\rs,\ls} = \overline{C}_{\rs', \ls'}~\overline{E}$, where the coin operator parameters have been rescaled as $\rs'=\alpha\rs$ and $\ls'=\beta\ls$. Intuitively, the Lorentz transformation rescales the fractional steps of the Clock QW by $\alpha$ (resp. $\beta$), while adding $\alpha-1$ (resp. $\beta-1$) more points to the lattice. In this way, the counter will reach $0$ just at the end of the patch, as it did before the transformation.

\subsection{Continuum limit of the Clock QW}

The Clock QW does not have a continuum limit because its coin operator is not the identity in the limit $\varepsilon \rightarrow 0$. However, by appropriately sampling the spacetime points, it is possible to take the continuum limit of a solution of the Clock QW and show that it converges to a solution of the Dirac equation, subject to a Lorentz transform with parameters $\rs$, $\ls$. Indeed, the limit can be obtained as follows. First, we divide the spacetime in lightlike rectangular patches of dimension $\rs\times\ls$. Second, we choose as representative value for each patch the point where the interaction is non-trivial, averaged according to the dimensions of the rectangle:
\begin{equation*}
\psi'(r,l) = \left( \begin{array}{c} \dfrac{\psi_+(\floor{r/\rs}_\varepsilon,\floor{l/\ls}_\varepsilon)}{\sqrt{\ls}} \\ \dfrac{\psi_-(\floor{r/\rs}_\varepsilon,\floor{l/\ls}_\varepsilon)}{\sqrt{\rs}} \end{array} \right).
\end{equation*}
Finally, by letting $\varepsilon\to 0$ we obtain
\begin{equation*}
\psi'(r,l) = S\psi(r/\rs,l/\ls)
\end{equation*}
where now the $r,l$ coordinates are to be intended as continuous.

Since $\psi'$ is of course a solution of the Dirac equation (with a rescaled mass), this proves that the continuum limit of the Clock QW evolution, interpreted as described above, is again the Dirac equation itself.

\subsection{Decoupling of the QW and the Klein-Gordon equation}
The Clock QW does not have a proper continuum limit unless we exclude the intermediate computational steps. Still, as we shall prove in this section, its decoupled form {\em has} a proper limit, which turns out to be the Klein-Gordon Equation with a rescaled mass. By a decoupled form, we mean the scalar evolution law satisfied by each component of a vector field, individually (see \cite{ArrighiKG}). In the following, we give the decoupled form of the Clock QW. The evolution matrix $W$ is sparse and allows for decoupling by simple algebraic manipulations, leading to:
\begin{equation}
  \left[T^{\ls+\rs} - a\tau^{-\ls}  T^\rs - d\tau^{\rs} T^\ls + \det(C) \tau^{\rs-\ls}\right]\psi = 0
\end{equation}
(where $T=e^{\varepsilon\partial_t}$ and $\tau=e^{\varepsilon\partial_x}$). This is a discrete evolution law which gives the value of the current step depending on three previous time steps, namely the ones at  $t=-\rs$, $t=-\ls$ and $t=-\rs-\ls$.

By expanding in $\varepsilon$ the displacement operators and assuming that the coin operator verifies:
\begin{equation}
\det(C) = 1, \quad a = d = 1 + \frac{\varepsilon^2m^2}{2} + O(\varepsilon^3)
\end{equation}
(which is the case if  $a=d=\cos(m \varepsilon)$)
we obtain the continuum limit:
\begin{equation}
  \left(\partial_t^2 - \partial_x^2 + \frac{m^2}{\rs \ls}\right)\psi = 0.
\end{equation}
Up to redefinition of the mass $m' = m/\sqrt{\rs \ls}$, this is the Klein-Gordon Equation. This reinforces the interpretation of the Clock QW as model for a relativistic particle of mass $m'$.

\section{The Clock Quantum Cellular Automata} \label{sec:ClockQCA}

One downside of the Clock QW is the fact that the dimension of the coin space varies according to the observer. Equipped with a formal, general notion of Lorentz transform and Lorentz covariance for QCA, we can now seek for an exactly covariant QCA of fixed, small, internal degree of freedom.

\subsection{From the Clock QW to the Clock QCA}

The idea of the Clock QW was to let the walker propagate during a number of steps to the right (resp. to the left), without spreading to the left (resp. to the right). In the absence of any other walker, this had to be performed with the help of an internal clock. In the context of QCA, however, the walker can be made to cross ``keep going'' signals instead.\\ 
The Clock QCA has wire dimension $d=3$, with orthonormal basis $\ket{q}$, $\ket{0}$, $\ket{1}$. Both $\ket{q}$ and $\ket{0}$ should be understood as vacuum states, but of slightly different natures as we shall see next. $\ket{1}$ should be understood as the presence of a particle.\\
Thus, the Clock QCA has scattering unitary a $9\times 9$ matrix $U$, which we can specify according to its action over the nine basis vectors. First we demand that the vacuum states be stable, i.e.
\begin{align*}
\ket{q}\otimes\ket{q}&\mapsto \ket{q}\otimes\ket{q},\\
\ket{q}\otimes\ket{0}&\mapsto \ket{0}\otimes\ket{q},\\
\ket{0}\otimes\ket{q}&\mapsto \ket{q}\otimes\ket{0},\\
\ket{0}\otimes\ket{0}&\mapsto \ket{0}\otimes\ket{0}.
\end{align*}
Second we demand that multiple walkers do not interact:
\begin{align*}
\ket{1}\otimes\ket{1}&\mapsto \ket{1}\otimes\ket{1}.
\end{align*}
Third we demand that the interaction between $\ket{1}$ and $\ket{q}$ be dictated by a massless Dirac QW, or ``Weyl QW'', i.e. the single walker goes straight ahead:
\begin{align*}
\ket{1}\otimes\ket{q}&\mapsto \ket{q}\otimes\ket{1},\\
\ket{q}\otimes\ket{1}&\mapsto \ket{1}\otimes\ket{q}.
\end{align*}
Last we demand that the interaction between $\ket{1}$ and $\ket{0}$ be dictated by:
\begin{align*}
\ket{1}\otimes\ket{0}&\mapsto a(\ket{0}\otimes\ket{1})+b(\ket{1}\otimes\ket{0}),\\
\ket{0}\otimes\ket{1}&\mapsto c(\ket{0}\otimes\ket{1})+ d(\ket{1}\otimes\ket{0}).
\end{align*}
This map is unitary provided that the $2\times 2$ matrix $C$ of coefficients $C_{11}=a$, $C_{12}=b$, $C_{21}=c$, $C_{22}=d$ is unitary. For instance we could choose, as for the Dirac QW, $a=d=\cos(m \varepsilon)$, $b=c=-\ii \sin(m \varepsilon)$. \\
The Clock QCA is covariant, even though its wire dimension is fixed and small, as we shall see.\\

\subsection{Covariance of the Clock QCA}

In order to give a precise meaning to the statement according to which the Clock QCA is covariant, we must specify our Lorentz transform. According to Section \ref{subsec:LorentztransformsQCA} we must provide a function $f$, which we take to be the identity, and an encoding  $E_\alpha: {\cal H}_d \longrightarrow {\cal H}^{\otimes\alpha}_d$ which we take to be:
$$
\ket{a}\mapsto \ket{a}\otimes \bigotimes_{\alpha-1} \ket{q}, 
$$
written from the bottom wire to the top wire as was the convention for QWs.
The intuition is that the $(\alpha-1)$ ancillary wires are just there to stretch out this lightlike direction, but given that $\ket{q}$ interacts with no one, this stretching will remain innocuous to the physics of the QCA.

Let us prove that things work as planned:  
\begin{widetext}
\begin{align*}
\overline {U} ~ \overline{E} (\ket{a}\otimes\ket{b}) &= \left(\prod_{i=0\ldots\alpha-1,j=0\ldots\beta-1} U_{m'}\right) (\ket{a}^{0,0}\otimes \bigotimes_{i=1\ldots\alpha} \ket{q}^{i,0}) \otimes (\ket{b}^{0,0}\otimes \bigotimes_{j=1\ldots\beta} \ket{q}^{0,j})\\
&= \left(\prod^{(i,j)\neq(0,0)}_{i=0\ldots\alpha-1,j=0\ldots\beta-1} U_{m'}\right) U(\ket{a}^{0,0}\otimes\ket{b}^{0,0})\otimes \bigotimes_{i=1\ldots\alpha} \ket{q}^{i,0} \otimes \bigotimes_{j=1\ldots\beta} \ket{q}^{0,j})\\
&= U(\ket{a}^{0,\beta-1}\otimes\ket{b}^{\alpha-1,0}) \otimes \bigotimes_{i=1\ldots\alpha} \ket{q}^{i,\beta-1} \otimes \bigotimes_{j=1\ldots\beta} \ket{q}^{\alpha-1,j}\\
&= \overline{E} U (\ket{a}\otimes\ket{b}).
\end{align*}
\end{widetext}
Hence, the Clock QCA is Lorentz covariant. Notice that things would have worked equally well if $E_{\alpha}$ had placed $\ket{a}$ differently amongst the $\ket{q\ldots}$. It could even have spread out $\ket{a}$ evenly across the different positions, in a way that is more akin to the Lorentz transform for the Dirac QW. 

\section{Conclusions} \label{sec:Conclusions}

In the context of QW and QCA, we have formalized a notion of discrete Lorentz transform of parameters $\alpha, \beta$, which consists in replacing each spacetime point with a lightlike $\alpha\times\beta$ rectangular spacetime patch, $\overline{C}_{m'}\overline{E}$, where $\overline{E}$ is an isometric encoding, and $\overline{C}_{m'}$ is the repeated application of the unitary interaction $C_{m'}$ throughout the patch (see Fig. \ref{fig:Lorentz transform}). We then formalized discrete Lorentz covariance as a form of commutativity: $\overline{E} C_m = \overline{C}_{m'} \overline{E}$. This commutation rule as well as the fact the $E$ is isometric can be expressed diagrammatically in terms of a few local, circuit equivalence rules (see Fig. \ref{fig:CovRules} and \ref{fig:OtherRules}), \`a la \cite{CoeckeBoxes}. This simple diagrammatic theory allows for non-homogeneous Lorentz transforms (Fig. \ref{fig:nonhomogeneoussimple}), which let you switch between non-inertial observers. Actually, it would be interesting to compare the respective powers of covariance under non-homogeneous Lorentz transformations versus general covariance under diffeomorphisms plus local Lorentz covariance.

First we considered the Dirac QW, a natural candidate, given that it has the Dirac equation as continuum limit, which is of course covariant. Unfortunately, we proved the Dirac QW to be covariant only up to first-order in the lattice spacing $\varepsilon$. This is inconvenient if $\varepsilon$ is considered a physically relevant quantity, i.e. if spacetime is really thought of as discrete. But if $\varepsilon$ is thought of as an infinitesimal, then the second-order failure of Lorentz-covariance is irrelevant. Thus, this result encourages us to take the view that $\varepsilon$ is akin to infinitesimals in non-standard analysis. Then, the Dirac QW would be understood as describing an infinitesimal time evolution, but in the same formalism as that of discrete time evolutions. As an alternative language to the Hamiltonian formalism, it has the advantage of sticking to local unitary interactions \cite{ArrighiJCSS}, and that of providing a quantum simulation algorithm.

Exact Lorentz covariance, however, is possible even for finite $\varepsilon$. This paper introduces the Clock QW, which achieves this property. However the effective dimension of its internal degree of freedom depends on the observer. Furthermore, the Clock QW does not admit a continuum limit, unless we appropriately sample the points of the lattice. Yet, its decoupled form does have a continuum limit, which is the Klein-Gordon Equation. It is interesting to see that there is a QW evolution which can be interpreted as a relativistic particle (since it satisfies the KG Equation), and yet not have a continuum limit for itself.

Finally, we introduced the Clock QCA, which is exactly covariant and has a three dimensional state space for its wires.

We leave the following question open: is there a systematic method which given a QW with coin operator $C$, decides whether it exists a Lorentz transform $E_{\alpha}$, $f_{\alpha,\beta}$ such that $\overline{E} C_m = \overline{C}_{m'} \overline{E}$, i.e. such that the QW is covariant? The same question applies to QCA; answering it would probably confirm the intuition that covariant QWs are scarce amongst QWs.

The simple theory presented here can be criticized on several grounds. First, one may wish for more explicit comparison with the continuum theory. This may be done along the lines of \cite{ArrighiDirac,DAriano}:  by letting the lattice spacing $\varepsilon$ go to zero, the convergence of the spacetime wavefunction solution of the Dirac QW can be shown to tend to the solution of the Dirac equation, in a manner which can be quantified. Second, one may argue that the very definition of the Lorentz transform should not depend on the QW under consideration. Similarly, one may argue that the transformed wave function should be a solution of the original QW, without modifications of its parameters. However, recall that $1+1$ dimensional, integral Lorentz transforms are trivial unless we introduce a global rescaling. Thus the discrete Lorentz transform of this paper may be thought of as a biased zooming in. In order to fill in the zoomed in region, one generally has to use the QW in a weakened, reparameterized manner.

On the one hand, this paper draws it inspiration from Quantum Information and a perspective for the future would be to discuss relativistic quantum information theory \cite{aharonov1984quantum, peres2004quantum} within this framework. On the other hand, it forms part of a general trend seeking to model quantum field theoretical phenomena via discrete dynamics. For now, little is known on how to build QCA models from first principles, which admit physically relevant Hamiltonians \cite{DAriano3D, elze2013action, farrelly2013discrete, t2013duality} as emergent. In this paper we have identified one such first principle, namely the Lorentz covariance symmetry. We plan on studying another fundamental symmetry, namely isotropy, thereby extending this work to higher dimensions.

\section*{Acknowledgements} The authors are indebted to A. Joye, D. Meyer, V. Nesme and A. Werner. This work has been funded by the ANR-10-JCJC-0208 CausaQ grant.

\bibliography{../Bibliography/biblio} 

\appendix

\section{First-order-only covariance of the Dirac QW}\label{App:FirstOrderOnlyProof}

\begin{figure*}
\includegraphics[width=\textwidth]{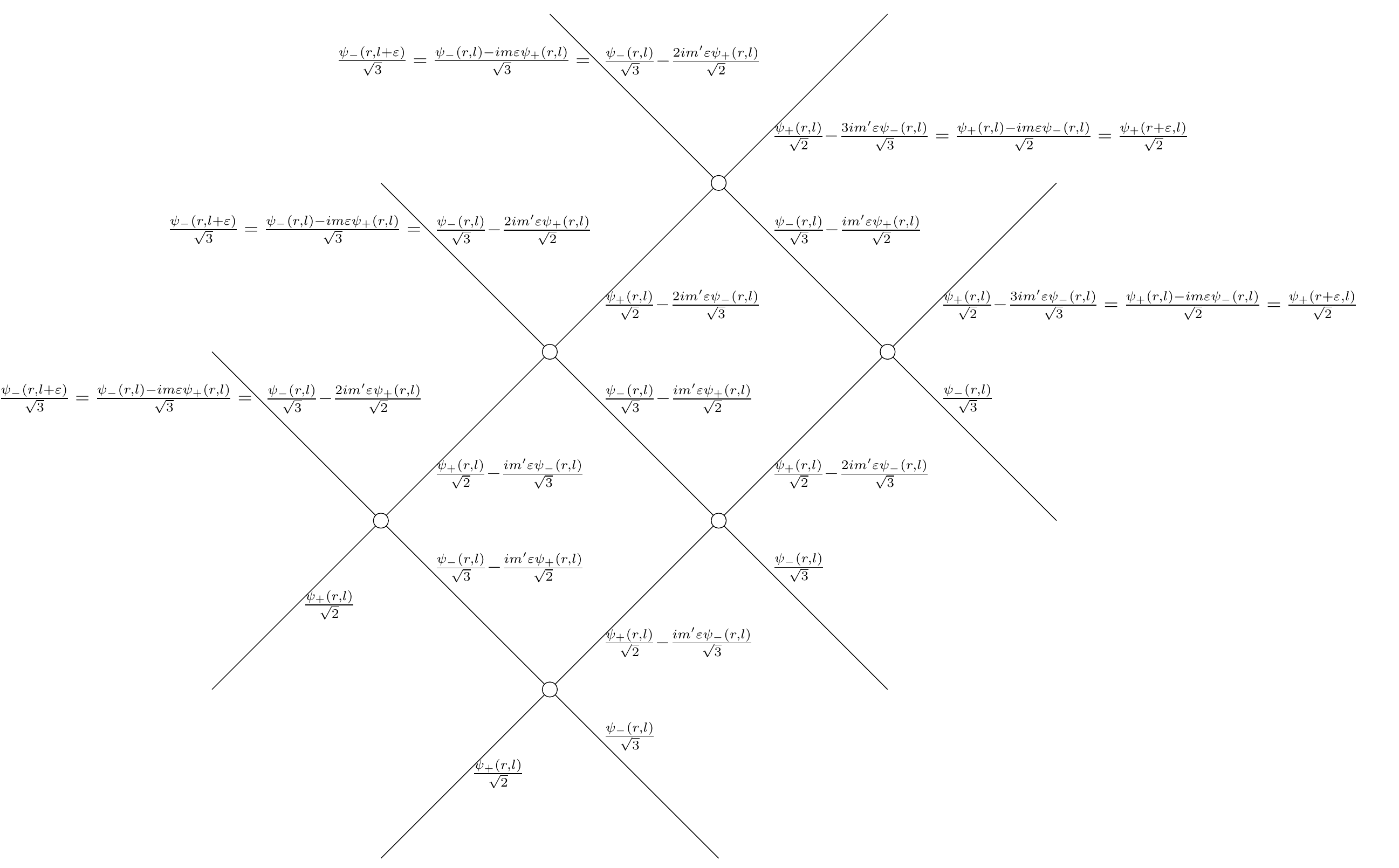}
\caption{\label{fig:FirstOrderCovariance} {\em First order covariance of the Dirac QW}. In the first order, the outcoming wires of a patch match the incoming wires of the next patch. Unless otherwise indicated, all the fields values appearing in this Figure are evaluated at $(r,l)$.}
\end{figure*}

{\em Uniqueness of encodings.} Here we prove that the only encoding compatible with first-order covariance is the flat one, as described in section \ref{subsec:DiracLorentzTransform}. In general, the encoding isometries $E_\alpha$, $E_\beta$ can be defined in terms of normalized vectors, $\mathbf v_\pm$ as follows (remember that for the Dirac QW, $\psi_+$ and $\psi_-$ are just scalars):
\begin{equation*}
E_\beta\psi_+ = \psi_+ \mathbf v_+, \qquad E_\alpha\psi_- = \psi_- \mathbf v_-.
\end{equation*}
In order to require covariance, we need to calculate the terms appearing in the commutation relation \eqref{eq:consistency}. The r.h.s of the relation is (see Fig. \ref{fig:FirstOrderCovariance} and Subsection \ref{subsec:focovariance}):
\begin{equation*}
\overline{C}_{m'}\overline{E} = \left(\begin{array}{c}
    \psi_+\mathbf v_+ - im'\varepsilon (\sum \mathbf v_-) \psi_- \mathbf 1_\beta \\
    \psi_-\mathbf v_- - im'\varepsilon (\sum \mathbf v_+) \psi_+ \mathbf 1_\alpha
  \end{array}\right) + O(\varepsilon^2)
\end{equation*}
where $\mathbf 1_d = (1,\dots,1)^{\mathsf T}$ is the $d$-dimensional uniform vector, and $\sum \mathbf v = \sum_i v_i$. On the other hand the l.h.s. is:
\begin{equation*}
\overline{E}C_m = \left(\begin{array}{c}
    \psi_+\mathbf v_+ - im\varepsilon\psi_-\mathbf v_+ \\
    \psi_-\mathbf v_- - im\varepsilon\psi_+\mathbf v_-
  \end{array}\right) + O(\varepsilon^2).
\end{equation*}
Requiring first-order covariance, one obtains
\begin{align*}
m\mathbf v_+ = m' \left(\sum\mathbf v_-\right) \mathbf 1_\beta, \qquad m\mathbf v_- = m' \left(\sum\mathbf v_+\right) \mathbf 1_\alpha
\end{align*}
which, together with the normalization of $\mathbf v_\pm$, gives
\begin{align*}
m' = \frac{m}{\sqrt{\alpha\beta}}, \quad \mathbf v_+ = \frac{e^{i\lambda_+}}{\sqrt{\beta}}\mathbf 1_\beta, \quad \mathbf v_- = \frac{e^{i\lambda_-}}{\sqrt{\alpha}}\mathbf 1_\alpha.
\end{align*}
thereby proving that the only possible encoding compatible with first-order covariance is the flat one (up to irrelevant phases).


\begin{figure}
\includegraphics[width=\columnwidth]{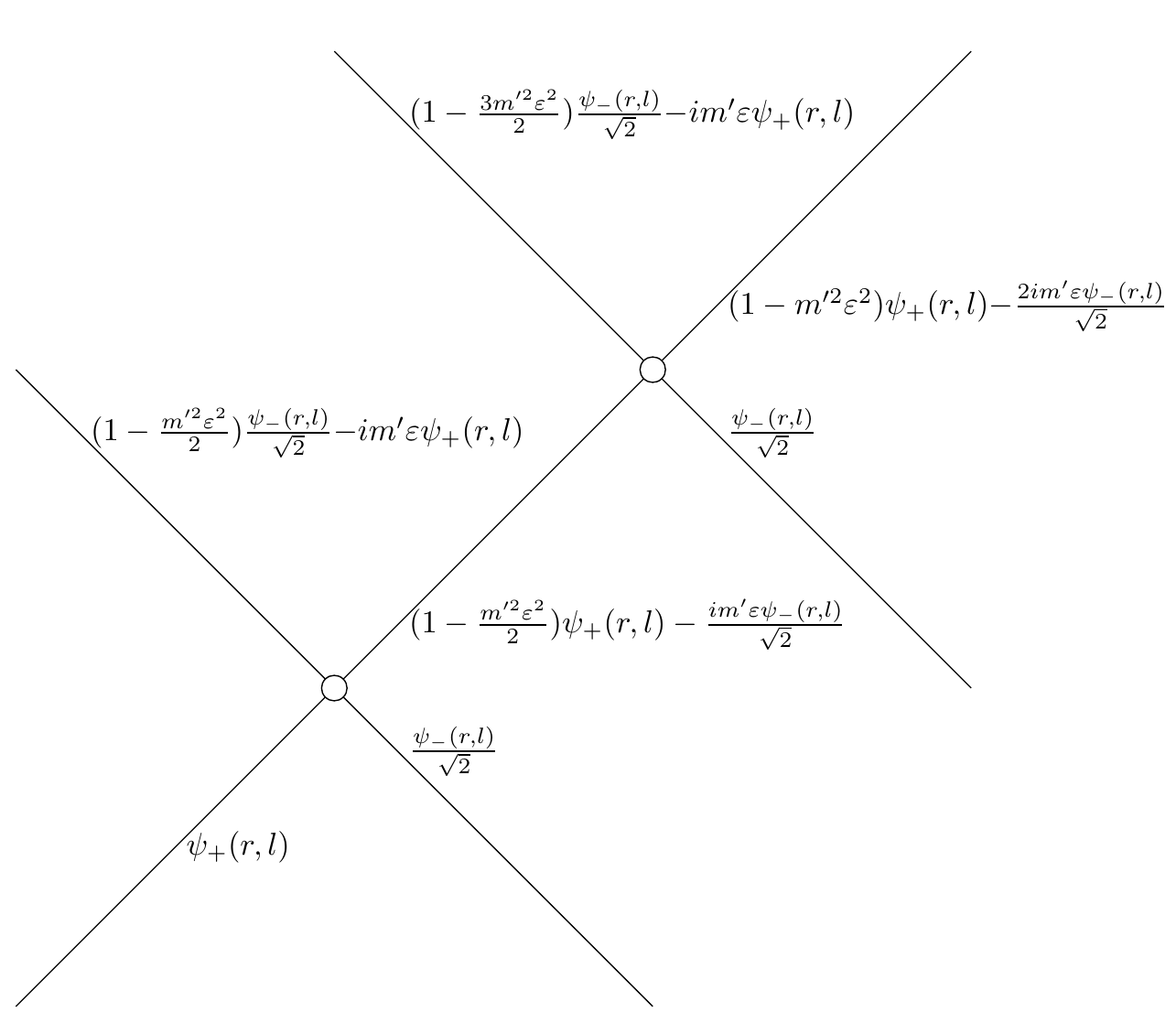}
\caption{\label{fig:FailsSecondOrder} {\em Failure of covariance at the second order for the Dirac QW.} The outcoming wires of a patch do not match the incoming wires of the next patch.}
\end{figure}

\medskip {\em Failure at second order.} The Dirac QW can similarly be expanded to the second order. This time, however, the patches that make up $\psi'$ do not match up. A simple counter-example supporting this fact arises with $\alpha=2$ and $\beta=1$ already, as illustrated in Fig. \ref{fig:FailsSecondOrder}. Notice that we ought to have $\overline{C}(0,1)\widecheck{\psi}_-(0,0)=\overline{C}(1,1)\widecheck{\psi}_-(0,0)$, if we want those outcoming wires to match up with the corresponding incoming wires of the next patch $\widecheck{\psi}_-(0,1)_0=\widecheck{\psi}_-(0,1)_1= \psi_-(0,1)/\sqrt{2}$. But it turns out that those outcoming wires verify $\overline{C}(0,1)\widecheck{\psi}_-(0,0)\neq \overline{C}(1,1)\widecheck{\psi}_-(0,0)$ due a term in $\varepsilon^2$.

\end{document}